\documentclass[acmsmall]{acmart}

\usepackage{hyperref}

\usepackage{hyperxmp}

\usepackage{amsmath,amsfonts}
\usepackage{algorithmic}
\usepackage{graphicx}
\usepackage{textcomp}
\usepackage{xcolor}
\usepackage{booktabs}
\definecolor{Gray}{gray}{0.9}
\usepackage{tabularx}
\usepackage{makecell}
\usepackage{verbatim}

\usepackage{multirow}
\usepackage[framemethod=TikZ]{mdframed}
\usepackage{xcolor}
\usepackage{tcolorbox}[boxsep=1mm]
\usepackage{array}

\usepackage[T1]{fontenc}
\usepackage[]{babel}
\usepackage[utf8]{inputenc}

\usepackage{chngcntr}
\counterwithin{figure}{section}

\usepackage{longtable}
\usepackage{lscape}
\usepackage{afterpage}
\usepackage{geometry}
\usepackage{pdflscape}

\mdfdefinestyle{implicationBox}{
    linecolor=blue,
    outerlinewidth=2pt,
    roundcorner=10pt,
    innertopmargin=10pt,
    innerbottommargin=10pt,
    innerrightmargin=10pt,
    innerleftmargin=10pt,
    backgroundcolor=gray!10!white
}

\UseRawInputEncoding

\AtBeginDocument{%
  \providecommand\BibTeX{{%
    \normalfont B\kern-0.5em{\scshape i\kern-0.25em b}\kern-0.8em\TeX}}}

\begin{document}

\title{The Psychological Costs of Artificial Intelligence Adoption in Software Engineering}

\author{Adam Alami}
\email{adal@cs.aau.dk}
\affiliation{%
  \institution{University of Southern Denmark}
  \streetaddress{The Maersk Mc-Kinney Moller Institute}
  \postcode{6400}  
  \city{S\o nderborg}
  \country{Denmark}
}

\author{Elda Paja}
\email{elpa@itu.dk}
\affiliation{%
  \institution{IT University of Copenhagen}
  \streetaddress{Rued Langgaards Vej 7}
  \city{2300 Copenhagen S}
  \country{Denmark}
}

\author{Abhishek Tiwari}
\email{abti@mmmi.sdu.dk}
\affiliation{%
 \institution{University of Southern Denmark}
  \streetaddress{The Maersk Mc-Kinney Moller Institute}
  \city{Odense}
  \country{Denmark}
}

\renewcommand{\shortauthors}{Alami, et al.}

\sloppy
\begin{abstract}

Artificial intelligence (AI) is increasingly used to augment software engineering (SE) workflows. While code generation remains the main use case, organizations are actively seeking AI integration in other practices such as test cases generation and code reviews. Organizational AI adoption strategies seem to focus on tangible outcomes such as productivity. However, AI is a disruptive force, introduced into settings where role identity, team norms, and the sources of job satisfaction were well established before the recent advances in generative AI. Historically, technological disruptions have caused psychological and social strains in workplaces, ranging from anxiety and eroded meaning to deskilling and disrupted professional identities. The assumption that AI for SE is cost-free may not be accurate. Therefore, in this study we sought to understand the psychological costs software professionals experience during organizational AI adoption. We carried out a case study in a large software development services company, one year after the company launched its AI adoption. We collected qualitative data through meetings and semi-structured interviews (N = 21). We found that software professionals experience \emph{accountability anxiety}, \emph{craft identity disruption}, \emph{meaning and satisfaction erosion}, \emph{cognitive and workload intensification}, and \emph{uncertainty distress}. Practitioners \emph{manage} these costs through practices that restore control, \emph{mitigate} them through protective and identity-preserving adaptations, or \emph{absorb} them, carrying what neither can resolve. We contribute to AI-human collaboration in SE by repositioning AI adoption as a human transition, not only a technological and organizational one.

\end{abstract}




\maketitle

\section{Introduction}
\label{sec:introduction}

\noindent Recent advances in generative Artificial Intelligence (AI) have produced commercial products such as Claude Code\footnote{\url{https://claude.com/claude-code}}, Codex\footnote{\url{https://openai.com/index/openai-codex/}}, and GitHub Copilot\footnote{\url{https://github.com/features/copilot/}}, which perform software engineering (SE) tasks traditionally carried out by humans~\cite{hou2024large}. Although these tools are most visibly associated with code generation, the Large Language Models (LLMs) underlying them are multi-task systems capable of supporting a broader range of SE activities, including those that have been traditionally entrusted to humans in a collaborative setting like code review~\cite{alami2025accountability,fan2023large}; for example, Claude Code had released a code review feature~\cite{anthropic2026codereview}\footnote{\url{https://code.claude.com/docs/en/code-review}}. The promise of productivity gains~\cite{ziegler2024measuring} across this widening span of tasks, amplified by intense industry discourse, is driving adoption in SE organizations~\cite{russo2024navigating,jensen2025managing}. For example, The Stack
Overflow Developer Survey 2025 (49,000$+$ respondents across 177 countries)
reports that 84\% of developers use or plan to use AI tools in their
development process, up from 76\% in 2024~\cite{stackoverflow2025ai}. according to the DORA report (``State of AI-assisted Software Development 2025''), tt the organizational level ``AI adoption has become nearly universal,'' 90\% of respondents use AI in their workflows and more than 80\% claim ``productivity'' increase\cite{dora2025}. 

However, this technological shift is assumed to carry no or few costs. History has taught us that such assumptions rarely hold. When mechanization transformed coal mining in the 1950s, Trist and Bamforth documented not only productivity disruptions but psychological consequences for the miners, including anxiety, loss of meaning, and the breakdown of role identities tied to the previous way
of working~\cite{trist1951some}. Braverman's analysis of automation argued that technologies introduced in the name of efficiency systematically degraded and deskilled the work of those who had mastered the displaced craft~\cite{braverman1998labor}. When computerization reached professional work, Zuboff showed that workers experienced the loss of action-centered skill as a crisis of competence and identity, even as the technology augmented what they could do~\cite{zuboff1988age}, and Barley's study of computed tomography in radiology demonstrated that a new technology reorders expertise, status, and professional self-understanding, not merely tasks~\cite{barley1986technology}.

Across these historical episodes, the pattern is consistent: technological transitions that reconfigure how work is done impose adaptation burdens on those who perform the work, and these burdens are psychological as much as they are workflow-related. There is little reason to assume that AI adoption in SE will be different. These historical parallels motivate our investigation of whether AI adoption generates comparable psychological consequences for software practitioners.

Early evidence on human-AI collaboration in SE suggests that many of the dynamics documented during previous technological transitions are already emerging in AI-assisted SE. AI appears to reshape how practitioners reason about responsibility, expertise, collaboration, and their own role in the development process. For example, Alami et al. reported that introducing LLM-assisted code review disrupts collective accountability by shifting responsibility to the individual instead of the collective level~\cite{alami2025accountability}. Subsequent studies further found that AI-generated reviews elicit different emotional, cognitive, and behavioural responses than reviews written by peers, altering how practitioners engage with feedback and collaborate during code review \cite{alami2025engagement,alami2025human}. Studies of AI-assisted programming report that developers continuously negotiate trust in AI-generated code, devote substantial effort to verification, and remain concerned about maintaining understanding and control over generated artifacts despite productivity gains~\cite{barke2023grounded,vaithilingam2022expectation}. Collectively, these findings suggest that the adoption of generative AI introduces changes that extend beyond task execution to the psychological and social experiences of software development.

While these studies demonstrate that AI adoption influences how software engineers collaborate, reason, and engage with their work, the evidence remains fragmented and largely detached from the organizational context in which adoption unfolds. Existing work explains the conditions shaping adoption, technology-related strain, and developer well-being, but not what practitioners psychologically incur as organizational AI adoption is enacted in software engineering practice. Therefore, we propose to investigate:

\medskip

\noindent \textbf{RQ:} \emph{What psychological costs do software practitioners experience during organizational AI adoption, and how do they manage these costs?}

\medskip

Drawing on the scope of psychological phenomena outlined by the American Psychological Association, encompassing cognitive, emotional, motivational, and social processes~\cite{apa}, and consistent with Lazarus and Folkman's view of ``stress'' as arising when individuals appraise environmental demands as exceeding their resources~\cite{lazarus1984stress}, we define psychological costs as the adverse cognitive, emotional, motivational, and social experiences that practitioners incur during organizational AI adoption. The term reflects how practitioners appraise and make sense of changes to their work, as distinct from the technical or organizational outcomes of adoption.

To answer our RQ, we carried out a single case study in a large Danish software development company one year into their AI adoption. We collected data through semi-structured interviews (n=21), member checking (n=12), and meetings (n=5), with various stakeholders; further methodological details in Sect.~\ref{sec:methods}.

In the studied case, AI adoption refers to the organizational introduction of commercial generative AI tools, coding assistants, agentic tools, and conversational LLMs, into software teams' workflows. Adoption was actively promoted. The organization deployed extensive internal messaging, designated practitioners as ``AI ambassadors'' to accelerate uptake, and set expectations through an ``AI maturity model'' against which practitioners assess their usage. We describe the adoption context in detail in Sect.~\ref{sec:methods}.

We found that, as AI adoption was enacted in software engineering practice, practitioners experienced recurring psychological costs that the organization's adoption strategy had neither anticipated nor addressed. Our analysis identified five distinct costs: \emph{accountability anxiety}, \emph{craft identity disruption}, \emph{meaning and satisfaction erosion}, \emph{cognitive and workload intensification}, and \emph{uncertainty distress}. Organizational conditions, notably adoption pressure and governance ambiguity, amplify these costs. Practitioners manage them through practices and adaptations ranging from deliberate oversight and skill-preservation routines to the reframing of professional identity, and, where these fall short, resigned adaptation.

Although we found little evidence that practitioners opposed AI adoption itself, we found that adoption created psychological costs. Variations in adoption primarily reflected compliance constraints, uneven applicability across roles and use cases, and tensions between experimentation and business-as-usual demands. Psychological costs therefore did not arise from resistance to AI; they emerged as practitioners attempted to integrate AI while preserving accountability, competence, professional agency, and meaning in their work.

We contribute:

\begin{itemize}

   \item[-] \textbf{How AI adoption becomes psychologically costly when enacted in SE practice.} Our contribution lies in explaining how psychological costs emerge as organizational AI adoption is enacted in SE. Our model (Fig.~\ref{fig:Model}) connects technological characteristics and organizational conditions to changes in practitioners' accountability, effort, identity, and meaning, and to the responses through which they accommodate these changes. This explanation provides a diagnostic framework for identifying where burdens originate, where organizational arrangements intensify them, and which aspects of engineering work may require intervention.\newline

   \item[-] \textbf{Repositioning AI adoption in SE as a human transition, not only as a technological and organizational one.} The AI adoption literature in SE has largely examined adoption through drivers, barriers, usability, and productivity outcomes. Our findings show that adoption is also, and simultaneously, a psychological transition in which practitioners' competence, identity, and relationship to their work are renegotiated. Historical accounts of earlier workplace technology shifts corroborate this pattern, but the contribution is to the SE adoption literature: models of AI adoption that omit the psychological dimension are incomplete, because the costs we document, may potentially shape adoption strategies.\newline
   
   \item[-] \textbf{Surfacing the unaccounted costs of AI adoption for adopting organizations.}  Practically, our study shifts the question from ``how do we accelerate AI adoption'' to ``what does adoption cost practitioners, and which organizational conditions intensify that cost.'' The amplifying conditions we identify, maturity-model pressure, efficiency expectations, and unresolved governance, are organizational choices and therefore actionable: organizations can clarify governance boundaries, decouple adoption expectations from individual measurement, and recognize verification work as legitimate effort. The costs rooted in the disruption itself cannot be prevented by better change management only, but they can be supported; the managerial buffering we observed indicates what such support looks like in practice.\newline

   \item[-] \textbf{Psychological costs cannot be reduced to resistance to AI.} Psychological costs coexist with acceptance of AI; they are consequences practitioners manage while adopting the technology, not evidence that they resist adopting it. We show that psychological costs arise among practitioners irrespective of their level of AI adoption (see Tbl.~\ref{tab:participants}, from exploratory to agentic use. Where adoption remained limited, this primarily reflected compliance constraints, uneven use-case applicability, and tensions with business-as-usual work rather than opposition to AI itself. 

\end{itemize}

\noindent In the remainder of this paper, we review related work in Sect.~\ref{sec:related}, then, we present methodological design in Sect.~\ref{sec:methods}. In Sect.~\ref{sec:findings}, we report the findings, discuss their implications in Sect.~\ref{sec:discussion} and trustworthiness in Sect.~\ref{sec:trust}, and conclude the paper in Sect.~\ref{sec:conclusion}.

\section{Related Work}
\label{sec:related}

The literature increasingly shows that generative AI does not simply automate SE. It redistributes engineering effort toward prompting, evaluation, correction, integration, and accountability. Yet most studies examine adoption at the individual level; not the assumptions built into organizational adoption strategies when those strategies encounter practice. We identified three streams of related works: \textbf{AI adoption in SE}, \textbf{Technostress}, and \textbf{well-being in GenAI}.

\subsection{AI adoption in SE}

Russo~\cite{russo2024navigating} provides a central account of the complexity of generative-AI adoption in SE. Its importance lies in moving beyond a narrow performance framing; adoption depends on how practitioners evaluate usefulness and fit, but also on trust, facilitating conditions, organizational support, and the surrounding work system. This makes adoption a relational accomplishment between the technology, engineering practices, and organizational arrangements, not a simple decision to deploy a tool~\cite{russo2024navigating}. The multiple-case reported by Kemell et al. extends this organizational view~\cite{kemell2025still}. The cross-case variation shows that organizations differ in maturity, constraints, objectives, risk tolerance, and implementation pathways. Such variation cautions against universal maturity models that imply a single linear route from experimentation to pervasive or agentic use~\cite{kemell2025still}.

Barke et al. show that programmers do not interact with code-generating systems in a single, stable mode~\cite{barke2023grounded}. Their work alternates between using AI for acceleration and using it as a source of exploration, with substantial attention devoted to interpreting suggestions and maintaining task control. This finding undermines the idea that generated code simply replaces human production effort~\cite{barke2023grounded}.

Zakharov et al. further show that developers attribute different roles to AI-powered development tools and that these role perceptions shape their adoption of the technology \cite{zakharov2025ai}. Whereas this work examines the roles practitioners assign to AI, we examine the corresponding changes practitioners experience in their own professional role and identity as AI assumes activities previously central to software production.

Vaithilingam et al. identify an expectation--experience gap; users may perceive value even when measured task outcomes do not show corresponding improvement~\cite{vaithilingam2022expectation}. This distinction is methodologically important. Perceived productivity is consequential, it shapes continued adoption, but it cannot be treated as equivalent to task performance, code quality, or organizational value~\cite{vaithilingam2022expectation}.

Liang et al.'s large-scale survey found that a primary reason developers decline AI coding assistants is the feeling of losing control over their code, alongside frustration with output that does not meet functional or quality expectations~\cite{Liang2024}.

Recent evidence sharpens the argument of the shift of SE towards verification. Fan et al. explicitly examine verification load and fatigue~\cite{fan2026help}, while Vella and Blincoe describe a longitudinal shift from creation toward ``supervisory engineering work'': directing, evaluating, and correcting AI output~\cite{vella2026impact}. Fortes et al. similarly show that AI-assisted development may enhance perceived productivity by offloading routine work, while also disrupting cognitive engagement and introducing interaction and verification demands \cite{fortes2026productivity}. Together, these studies show that engineering effort is redistributed rather than displaced. We extend this evidence by explaining how verification becomes psychologically costly when practitioners must reconstruct sufficient understanding to remain accountable for AI-generated work.

Existing research explains what shapes AI adoption and how developers use and experience AI. We extend this literature by examining organizational AI adoption from the standpoint of practitioners who enact it in software engineering practice. Our case study reveals what adoption demands from practitioners and how its consequences reshape their work, roles, and professional identities. Our conceptual lens repositions AI adoption from an uptake outcome to a human transition whose costs are carried by the practitioners who enact it in SE reality.

\subsection{Technostress}

Technostress research examines stress created by technology-related demands such as overload, complexity, uncertainty, intrusion, insecurity, and continual change~\cite{ayyagari2011technostress,mubarkoot2026workarounds}. GenAI can operate simultaneously as a resource and a demand. It can reduce monotony, accelerate information access, support ideation, and increase perceived competence.

The technostress tradition does not imply that technology is inherently stressful~\cite{tarafdar2007impact}. Stress emerges through the relationship between environmental demands, personal and collective resources, and appraisal~\cite{ayyagari2011technostress,tarafdar2007impact,ragu2008consequences,jeyam2026still}. The same GenAI capability may be experienced as a challenge that enables mastery, a hindrance that obstructs valued work, or a threat to competence, control, or employment.

Classic technostress creators remain relevant but take new forms. Techno-overload appears when AI increases expected throughput or creates more output to inspect. Techno-complexity concerns the expertise required to prompt, evaluate, integrate, and govern probabilistic outputs. Techno-uncertainty~\cite{ayyagari2011technostress,tarafdar2007impact} intensifies because models, interfaces, compliance, and capabilities change rapidly. Techno-insecurity~\cite{ragu2008consequences} includes fears of replacement, devaluation, or falling behind. Techno-invasion~\cite{ayyagari2011technostress,ragu2008consequences} arises when continuous learning and experimentation extend beyond ordinary work boundaries.
More recently, Kwon et al. identified seven AI-induced technostressors across professional settings and documented the coping strategies professionals use in response \cite{kwon2026investigating}. This work demonstrates that AI introduces stressors and coping needs beyond established technostress categories; however, it remains occupation-general and organizes the phenomenon around stressors and coping, rather than examining how organizational adoption reshapes the accountability, agency, identity, and meaning attached to software engineering work.

Our construct of psychological costs builds on both established technostress research and emerging work on AI-induced stressors and coping, but differs in its analytical focus. %
Technostress primarily explains how technology-related demands are appraised and translated into strain. Psychological costs instead capture what practitioners expend, lose, or place at risk while adapting to AI-mediated changes in their SE work. In the context of SE, these costs include not only stress arising from technological demands, but also intensified verification effort, accountability anxiety, displaced professional agency, disrupted craft identity, and eroded meaning. They may therefore arise across levels of AI adoption, from exploratory to agentic use, and coexist with acceptance, enthusiasm, and perceived productivity benefits. 

Technostress is intentionally technology, and occupation-general; it explains demands that can arise across users, technologies, and work settings. Our conceptual lens is deliberately specific to SE. Generative AI does not merely introduce another demanding technology into software work; it assumes activities through which practitioners exercise expertise, authorship, and professional agency. AI is potentially disruptive not only to work processes but also to professional identity. Psychological costs therefore explain what software practitioners expend, lose, or struggle to preserve as AI participates in producing the artifacts for which they remain professionally responsible. Although there are conceptual parallels between our findings and technostress, our contribution lies in explaining how AI technology, engineering obligations, and organizational adoption conditions interact to produce psychological costs in SE, providing a basis for identifying where organizational intervention may be directed (see Sect.~\ref{sec:discussion} for further discussion).

\subsection{Well-being in GenAI in SE Era}

Well-being research provides a broader lens than technostress for examining how GenAI affects practitioners. Workplace well-being encompasses not only the absence of strain, but also positive functioning, satisfaction, engagement, autonomy, competence, relatedness, professional growth, and meaningful work~\cite{ryff1989happiness,ryan2000self,bakker2007job}. This distinction is important because practitioners may avoid stress while still experiencing a loss of agency, professional fulfillment, or connection to their work.

Emerging SE research questions the field's dominant emphasis on whether GenAI improves developer productivity. Guizani et al. argue that this narrow framing can conceal consequences for developers' cognitive load, stress, burnout, work--life balance, and professional development~\cite{guizani2026cost}. They call for GenAI research to evaluate sustainable productivity; whether productivity gains can be maintained without undermining the people responsible for producing and assuring software.

Brandebusemeyer et al. provide empirical support for moving beyond productivity assessment~\cite{brandebusemeyer2026developers}. Their mixed-methods field study shows that developers generally value GenAI for monotonous, repetitive, and structured tasks and perceive improvements in efficiency and workload. However, these benefits depend on the task, interaction mode, and quality of the generated output. AI interaction can itself create cognitive load, while combining code suggestions and chat-based interaction within the same task may diminish their benefits. Developer experience is therefore shaped not simply by whether GenAI is used, but by how its capabilities fit the work being performed~\cite{brandebusemeyer2026developers}.

These studies establish GenAI as a potential resource and a potential demand in SE. It may reduce repetitive effort, overcome task blockages, and improve satisfaction. At the same time, it can introduce oversight labor, verification fatigue, continuous-learning pressure, and escalating expectations regarding output and pace~\cite{guizani2026cost,brandebusemeyer2026developers}. Productivity gains and well-being gains must therefore be examined as related but distinct outcomes.

These streams of work explain important but separate parts of the phenomenon. AI-adoption research identifies the conditions shaping uptake and use; technostress explains how technology-related demands become strain; and well-being research evaluates whether AI-supported work remains psychologically sustainable. What remains insufficiently explained is what software practitioners incur as an organizational adoption strategy is enacted through SE practice. Therefore, psychological costs identify the intensified effort, accountability anxiety, disrupted agency, threatened craft identity, and erosion of meaning that AI adoption may produce. These costs may remain concealed when adoption is evaluated through usage, acceptance, or productivity, even when developers remain satisfied with GenAI and recognize its benefits.

Our contribution is therefore not another account of whether developers accept AI or whether AI impacts negatively or positively their well-being, but an explanation of what organizational AI adoption demands from software practitioners, why those demands arise in SE work, and how practitioners carry them.

\section{Methods}
\label{sec:methods}

\noindent We adopted a holistic single case study design~\cite{yin2018case}. The case is a large Danish software development company, approximately one year into an organization-wide AI adoption at the time of the study. Our rationale for a single case design follows Yin's ``common case'' logic~\cite{yin2018case}; the objective is to capture the circumstances and conditions of an everyday situation, here, an organization undergoing AI adoption, because of the insights it can provide into the phenomenon as it unfolds across the software industry. The case also offered a unique empirical access; the organization permitted inquiry into the lived experience of its practitioners during, rather than after, the transition, allowing us to study psychological costs and practitioners' responses as they were being experienced and formed, not retrospectively reconstructed~\cite{yin2018case}. The single case serves to develop and ground constructs and mechanisms that can be examined in other settings, consistent with case study research traditions~\cite{yin2018case,runeson2009guidelines}.

\subsection{Case Selection and Recruitment}
\label{sec:case-selection}

The case was recruited through AI4SE\footnote{\url{https://portal.findresearcher.sdu.dk/da/projects/human-centered-adoption-of-artificial-intelligence-for-software-e/}}, a Danish research project on AI adoption in SE, in which several companies partner with the University of Southern Denmark (SDU) and other research institutions. Recruitment occurred during the formation of the project consortium. This constellation aligned the interests of both parties; the organization sought insights into its adoption effort, while our objective was to study the practitioner experience of AI adoption as it unfolds. The company joined the project, in part, to gain knowledge from the experiences of other consortium participants and contributes with its own experience.

\subsection{Case Description}
\label{sec:case-description}

The case, hereafter \textsc{SoftHouse} (a pseudonym), is a large, long-established Danish software development company. \textsc{SoftHouse} develops, maintains, and continuously evolves software products for critical and highly regulated domains, alongside offering bespoke software development services. Several of its products are well-established systems underpinning services in public health, government administration, and taxation; some have been in operation for over two decades and undergo continuous modernization efforts. The company's products are operating in over 50 countries. The company also continuously seeks to expand into new domains, such as asset and public infrastructure security. Development effort is typically concentrated on evolving existing products; implementing new features upon customer requests, or innovating products in response to market demands and shifts.

\textsc{SoftHouse} operates in markets where reliability, security, and regulatory compliance are contractual and legal obligations rather than aspirations; its customers include public institutions and private organizations for whom software failure carries safety, legal, or societal consequences. This positioning shapes \textsc{SoftHouse}'s engineering culture; quality assurance, traceability, and accountability for delivered software are deeply institutionalized, enforced through a
multi-layered compliance regime combining external legislation governing data, certifications governing process maturity, and internal audits governing output conformance.

Organizationally, \textsc{SoftHouse} is structured into ``business units,'' each specialized in a specific business domain (e.g., health, taxation). Each business unit is, in turn, organized into projects mandated to deliver specific scopes, ranging from feature development on existing products to the development of entirely new products. \textsc{SoftHouse}'s workforce comprises 1,200 employees, the majority of whom are
software practitioners; software engineers (50\% of its workforce), testers and test managers, architects, product owners, project managers, and technical leads.

\textsc{SoftHouse} is also a proud agile organization. It rolled out Scrum across its development over two decades ago, and agile ways of working are deeply embedded in how teams plan, deliver, and collaborate. The AI adoption examined in this study thus entered an organization with mature, long-institutionalized development practices
and a strong sense of how software should be built.

\subsubsection*{\textbf{AI Adoption and Its Context}}
\label{sec:adoption-context}

\textsc{SoftHouse} launched an organization-wide AI adoption initiative in January 2025, approximately one year before our study started. The initiative introduced commercial generative AI tools into practitioners' workflows. The first product rolled out at the enterprise level was GitHub Copilot, which was abandoned three months later due to what the organization described as ``quality issues;'' software engineers reported ``code quality issues'' and an ``overall dissatisfaction with the product.'' It was replaced by Claude Code and Claude Desktop.

The adoption was supported rather than prescribed. The integration of AI into SE
workflows was not mandated; projects were encouraged to experiment and
share what they learned, locally and at the enterprise level. \textsc{SoftHouse} established a dedicated AI program with enterprise-wide oversight of the adoption, led by an appointed program director. Each project was assigned \emph{AI ambassador}, practitioners whose role was to promote AI adoption and facilitate knowledge sharing within and across projects. However, individuals assigned to these roles did not necessarily have prior experience with AI or its adoption into SE. The organization further set up a dedicated knowledge hub to consolidate and disseminate learning, and deployed extensive internal messaging; participants described AI as a recurring subject of company briefings and meetings. It also introduced an AI maturity model defining levels (i.e., 1--5, novice to agentic-AI) of AI-assisted practice, from novice use to advanced agentic use, against which practitioners assess their own usage, with organizational ambitions for practitioners to reach the upper levels.

At the time of the study, the internal governance of AI use was still taking form. Rules governed data classification, online access, and the delegation of credentials to agentic tools, while a comprehensive internal AI policy remained under development. Permitted AI use also varied across projects, as customer and domain requirements imposed constraints in some engagements; for example, prohibition of commercial cloud-based tools in some client's projects. Practitioners therefore encountered the adoption under heterogeneous conditions; organizationally championed and infrastructurally supported, yet not prescribed, and bounded by evolving internal rules and project-specific restrictions, either driven by domain-specific compliance or customer's requirements on how to use AI. One year in, the initiative had not met the organization's original expectations, i.e., ``achieving 50\% adoption rate among software engineers,'' prompting \textsc{SoftHouse} to seek avenues to reinvigorate its adoption strategy, including its participation in the research project through which this study was conducted.

\subsubsection*{\textbf{Experiment-and-Share as an AI Adoption Strategy}}
\label{sec:experiment-share}

\textsc{SoftHouse} did not define or prescribe how AI should be integrated into SE workflows, instead, they pursued what we characterize as an \emph{experiment-and-share} strategy; practitioners and projects were encouraged to experiment with AI within their respective compliance boundaries, build competence through use, and share what they learned, locally within projects and across the enterprise. A senior manager (Tbl.~\ref{tab:participants} documents our sample) described the approach as deliberately phased: \emph{``we throughout phases has been trying to both experiment, to build competence, to also assess maturity''} (P13). The AI maturity model served as the assessment instrument within this strategy, while the AI accelerators and the knowledge hub constituted its diffusion mechanisms; experimentation was expected to happen within projects, and its lessons to be consolidated centrally and shared across the enterprise. In some settings, this took organized forms; participants described working groups tasked with determining \emph{``how do we actually do this, how are we going to work going forward with these tools''} (P14).

Two considerations underpinned this strategy. First, the heterogeneity of \textsc{SoftHouse}'s compliance landscape made uniform prescription impractical; what is permitted varies by domain, customer, and product, and the organization must remain, in P13's words, \emph{``very specific on where do we apply and how much are we allowed to apply in these different domains''} (P13). In the most sensitive domains, AI use is categorically restricted or confined to isolated configurations, such as locally hosted models. Second, the strategy reflected a leadership stance that combined conviction about AI's gains with a commitment to human oversight: \emph{``we very much believe in keeping humans in the loop ... but on the other hand we really also foresee that there is productivity and efficiency gains with the use of AI''} (P13). Expected gains were also tied to the organization's growth trajectory; leadership anticipated that AI-enabled productivity would allow the organization to scale differently than in the past.

The strategy, however, operated within a field of competing external expectations. While compliance obligations demanded restraint, customers, partners, and competitors pressed for speed; P13 described the resulting position: \emph{``we need to be aligned with compliance and security, but on the other hand we have competitors, we have customers, we have partners saying, well, why don't you guys just give it full speed''} (P13). The experiment-and-share strategy was, in this sense, the organization's mechanism for advancing adoption under conditions in which neither mandating nor prohibiting AI use was available to it.

The interviews' data show mixed results. From an adoption perspective, some practitioners achieved agentic level adoption (e.g., P7 and P14). Others (e.g., P19) met rigid compliance requirements, allowing them to use only locally hosted models; the experiment failed due to \emph{``limited computation power''} (P19). Knowledge sharing also experienced mixed results. However, where adoption remained limited seems to be primarily a reflection of compliance constraints, uneven use-case applicability, and tensions with business-as-usual work rather than opposition to AI itself (see sect.~\ref{sec:findings}).

Knowledge sharing also experienced mixed results. Where sharing was locally scaffolded, it functioned; participants described front runners mentoring colleagues through team-level adoption (P16), and the accelerators relaying use cases and reassurance across projects (P11). Elsewhere, the strategy's sharing premise met friction of three kinds. First, what was shared did not always transfer across
roles; as a test engineer explained, \emph{``the ones[experiments] that were shared across a lot of us weren't really relevant for my role''} (P10), leaving practitioners outside the development track without usable guidance. Second, sharing was socially costly; some practitioners withheld their experiences for fear of judgment, perceiving a high barrier to exposing their AI practices to colleagues (P1). Third, governance ambiguity suppressed sharing at its source; practitioners hesitated to disclose experiments conducted in unclear regulatory territory, \emph{``I am very hesitant at sharing this''} (P14), precisely the experiences the strategy depended on circulating. The strategy's central premise, that locally discovered lessons would be shared, thus held unevenly; knowledge flowed where roles were similar, psychological safety was
sufficient, and permissibility was clear, and stalled where any of the three was absent.

\subsection{Participants and Data Collection}
\label{sec:data-collection}

\begin{figure*}[!t]
    \includegraphics*[trim=0.5cm 4cm 0.5cm 1cm, clip, width=1.0\textwidth]{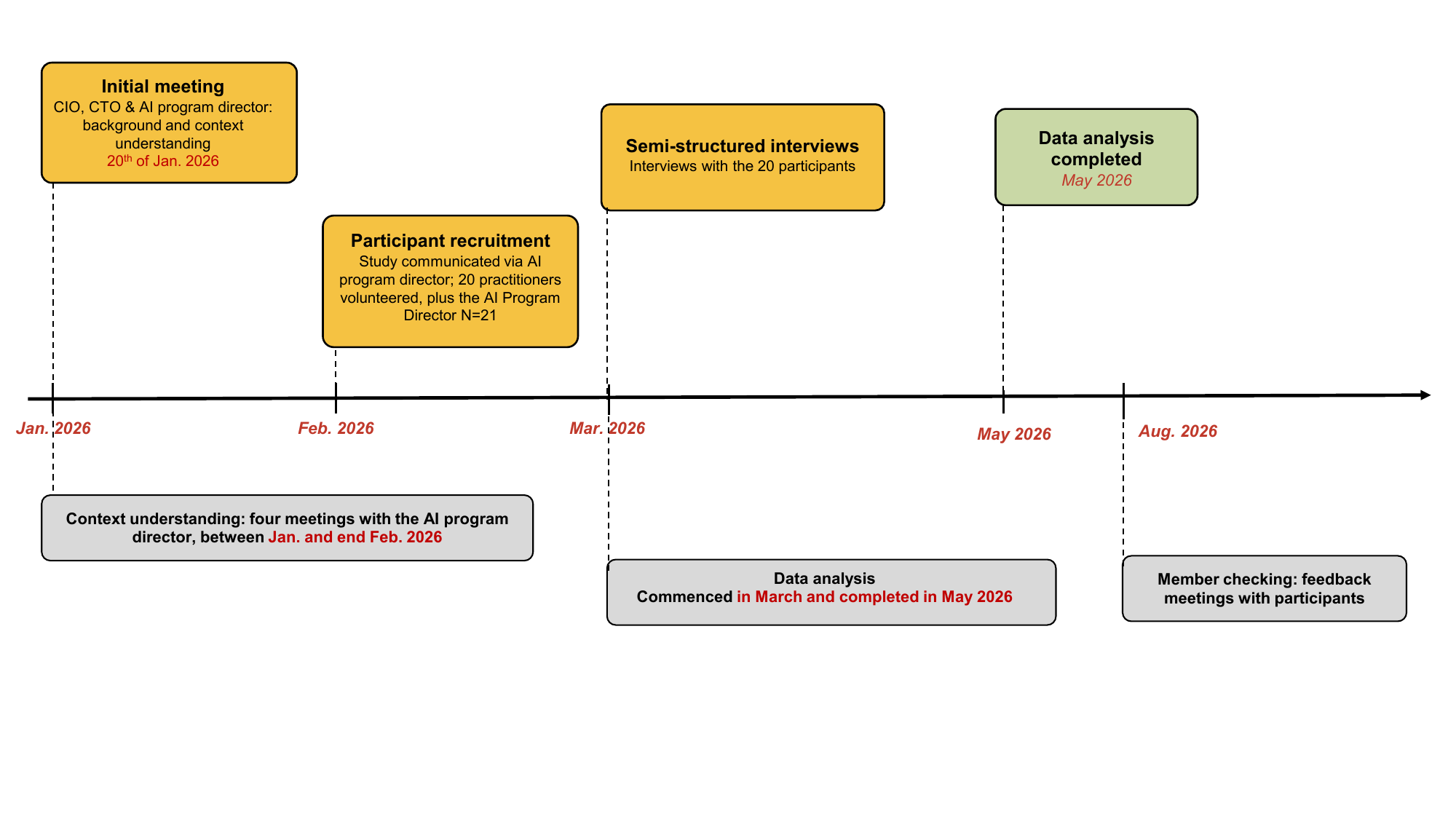}
        \caption{Chronological overview of the data collection and analysis process (January 2026--August 2026).}
        \label{fig:Timeline}
        
\end{figure*}

\begin{table}[!th]
\caption{Overview of stakeholder meetings held for background and context understanding.}

\label{tab:meetings}
\renewcommand{\arraystretch}{1.0}
\footnotesize

\begin{tabular}{lp{5.5cm}p{4cm}ll}
\toprule
\textbf{ID} & \textbf{Meeting's purpose} & \textbf{Attendees} & \textbf{Date} & \textbf{Duration} \\
\midrule

M1 & Background and context understanding & CTO, CIO, and AI program director & Jan. 2026 & One hour\\
\addlinespace
M2 & AI adoption program details (follow-up 1) & AI program director & Jan. 2026 & 30 mins\\ 
M3 & AI adoption program details (follow-up 2) & AI program director & Feb. 2026 & 30 mins\\
M4 & Adoption progress: further details (follow-up 3) & AI program director & Feb. 2026 & 30 mins\\
M5 & Adoption progress: further details (follow-up 4) & AI program director & Feb. 2026 & 30 mins\\
\bottomrule

\end{tabular}
\end{table}

We collected data through two complementary means; Fig.~\ref{fig:Timeline} documents the data collections and analysis timeline. First, to build an understanding of the adoption context, we held meetings with key stakeholders, including the AI program director and other senior managers. These meetings informed our reconstruction of the adoption's history, structure, and governance (Sect.~\ref{sec:adoption-context}) and were documented through audio recording and transcription. Second, to capture practitioners' experiences of the adoption, we conducted semi-structured interviews with practitioners across roles (Tbl.~\ref{tab:participants}).

Participants were recruited through the AI program director. She approached project managers, communicated the study and its intent, and relayed our call for participants; twenty one practitioners volunteered to participate, including the AI program director. The participants were known to the AI program director and first author. Although managers communicated the recruitment, practitioners were asked to contact the AI program director directly. Therefore, participants are anonymous to the organization except the AI program director and the first author.

We sought a population representative of the organization's practitioner composition across roles and experience levels. Our sample reflects this diversity; half of the participants are software engineers, with the remainder spanning testing, project management, architecture, and technical leadership roles. Table~\ref{tab:participants} documents the participants, their roles, and tenure. Participation was voluntary, and participants were assured anonymity toward both the research audience and their own organization; no information on who participated, declined, or withdrew was shared back to the organization. The data collected was not shared with any party in the organization; only the final results (see Sect.~\ref{sec:member-checking}).

The resulting sample reflects the practitioner composition of \textsc{SoftHouse}'s delivery organization. Almost half of the participants (10/21) are software engineers spanning seniority levels from junior to lead, complemented by architects (3), testing roles (3), senior program managers (2), AI program director, and heads of business units (2), providing perspectives from both the practitioner and delivery-leadership levels. The sample comprises five women and sixteen men. Participants' tenure at \textsc{SoftHouse} ranges from 1 to 23 years (median 9.5), predominantly with practitioners socialized into the organization's engineering practices long before the AI adoption. Their professional experience in the software industry ranges from two to 40 years (median 20); nineteen of the 21 participants have a decade or more in the industry, meaning the sample is dominated by practitioners whose professional identities and skills were formed well before the emergence of generative AI, the population for whom the transition under study carries the highest stakes (see Sect.~\ref{sec:trust} for threats to validity discussion).

All interviews were conducted online using MS Teams or in person and in English. All meetings and interviews were audio recorded, then transcribed. Interviews lasted on average 50 minutes. Although we designed an interview guide, its purpose was to steer the conversation while allowing fluidity, prompting interviewees to elaborate based on their responses. Table~\ref{tab:guide} documents key parts of the interview guide. The interviews generated approximately 290 pages (in average, 14 pages each) of transcription and 17 hours and 40 minutes of audio. Table~\ref{tab:meetings} documents our meetings. Data collection, including transcriptions and managing the relationship with \textsc{SoftHouse} were carried out by the first author.

\begin{table*}[!t]
\caption{Interview participants. ``M'' refers to man and ``W'' to woman. ``SE'' refers to Software Engineers, ``BU'' to Business Unit, and ``PM'' to Program Manager. ``Tenure'' refers to years in \textsc{SoftHouse} at the time the AI adoption was initiated; ``Exp.'' refers to total years of professional experience in the software industry. (*) indicates participants who took part of member checking.}

\label{tab:participants}
\renewcommand{\arraystretch}{0.80}
\footnotesize

\setlength{\tabcolsep}{4pt}
\begin{tabular*}{\textwidth}{@{\extracolsep{\fill}}lp{1.5cm}ccp{3.0cm}p{3.5cm}c@{}}
\toprule

\textbf{ID} & \textbf{Role} & \textbf{Tenure} & \textbf{Exp.} & \textbf{AI Tools} & \textbf{SE Activities} & \textbf{Adoption Level}$^{\mathrm{a}}$ \\
\midrule

\addlinespace[4pt]

P1(W)  & Test Manager& 10 & 25 & Claude Desktop & Testing, test management & IA \\

P2(M)  & Senior SE & 5  & 10 & Claude Code & Coding, code review & DO \\

P3(M)  & Senior SE & 10 & 20 & Claude Code & Coding, testing, requirements tracing & AG \\

P4(M)  & Lead Architect & 10 & 40 & Claude Desktop & Test design, requirements \& release documentation & IA \\

P5(M)  & Lead Architect & 10 & 25 & Claude Desktop, Claude Code & Architecture, adoption facilitation & DO \\

P6(M)(*)  & Lead SE & 7  & 15 & Claude Code & Process automation & IA \\

P7(M)(*)  & Lead SE & 10 & 12 & Claude Code, CodeScene & Coding, architecture, code review & AG \\

P8(W)(*)  & Head of BU & 8  & 26 & Claude Desktop & Content generation & IA \\

P9(M)(*)  & Test Engineer& 7  & 16 & Claude Desktop, Claude Code & Testing, process experimentation & DO \\

P10(W) & Test Engineer & 5  & 10 & Claude Desktop, Claude Code & Testing & IA \\

P11(M) & Senior PM & 12 & 40 & Claude Desktop, Claude Code & Project management, documentation & IA \\

P12(M)(*) & Senior SE & 3  & 10 & Local LLM, Claude Code & Coding, code review, scripting & CF \\

P13(M)(*) & Head of BU & 23 & 35 & Claude Desktop & Documents generation & IA \\

P14(M)(*) & SE & 12 & 12 & Claude Code, custom agents & Coding, debugging, testing, agent development & AG \\

P15(W)(*) & Senior PM & 5  & 30 & Claude Desktop & Document generation & IA \\

P16(M)(*) & Senior SE & 10 & 14 & Claude Desktop, Claude Code & Coding, code review, team leadership & IA \\

P17(M) & Principal Architect & 10 & 30 & Claude Desktop & Architecture, prototyping, documentation & DO \\

P18(M)(*) & Lead SE  & 9  & 20 & Claude Code & Coding, performance optimization, code analysis & DO \\

P19(M) & Lead SE & 8  & 30 & Claude Code & Platform engineering, coding, code review & EX \\

P20(M)(*) & Junior SE & 1  & 2  & Claude Code & Full-stack coding, prototyping & IA \\

P21(W)(*) & AI Program Director & 12  & 35  & Claude Desktop & Document generation & IA \\

\addlinespace[4pt]

\toprule

\addlinespace[4pt]

\multicolumn{7}{@{}p{\dimexpr\textwidth-2\tabcolsep\relax}@{}}{$^{\mathrm{a}}$Adoption levels, as described in the interviews: \textbf{IA} = integrated assisted use (routine assistance inside daily work, such as completion, drafting, and structured prompting; the practitioner still authors most output); \textbf{DO} = delegated use with human oversight (substantial tasks handed to AI tools, governed by plan review, and verification practices); \textbf{AG} = agentic with human-in-the-loop (automated multi-step workflows with human gates); \textbf{EX} = Exploratory, but constrained, adoption is reduced by compliance, and client requirements.} \\

\end{tabular*}
\end{table*}

\begin{table}[!th]
\caption{Key parts of the interview guide relevant to this study's RQ.}

\label{tab:guide}
\footnotesize

\begin{tabular}{p{13cm}}
\toprule

\textbf{Introduction} \\
\midrule

Can you please introduce yourself? (Include educational background,
current role in the company, and years of experience in software
development.) \\
How did you experience AI adoption at \textsc{SoftHouse}?\\
How did you become involved with AI in your work? What were your initial perceptions when the AI adoption was introduced? \\

\midrule

\textbf{Section I: Experiencing the AI adoption} \\
\midrule

Can you take me through how you use AI in your workflows today? How did your usage evolve since the adoption started? \\
How do you feel about this change to your work and your role? \\
How are you copping with the change, whether technical or non-technical?\\
Do you have any concerns related to AI in your work? (Probed across
levels: personal, professional, and product/deliverable.) \\

\emph{Probing questions:} Probing focused on the felt experience behind reported practices and concerns. Examples: \\
   
    \quad What is behind that concern? Is it about the quality of your work, your professional reputation, or your job? \\
    \quad How did that make you feel? \\
    \quad How do you experience being responsible for code you did not write yourself\\
    \quad How do others around you (colleagues, your team) experience this change? \\
    \quad How this particular concern arise? For example, is it your own feelings, your peers' judgment, or how the organization does things?\\

\midrule

\textbf{Section II: Navigating the adoption} \\
\midrule

How do you stay in control of what the AI produces? How do you avoid becoming complacent? \\
How do you manage within the constraints that apply to your project (e.g., security, compliance, client restrictions)? \\
Have concerns like these been raised with you, or by you, in your team? How were they addressed? \\

\emph{Probing questions:} Probing focused on concrete practices, their rationale, and their perceived effectiveness. Examples: \\

    \quad Can you walk me through how you review or verify AI output? \\
    \quad Why do you do it that way? What would happen if you did not? \\
    \quad Do you do anything to maintain your own skills? \\
    \quad To what extent do these practices resolve the concern for you? \\

\midrule

\textbf{Conclusion} \\
\midrule

Based on your experience of this AI adoption, how would you summarize the main takeaways? \\
Is there anything else you would like to share about your experience? \\

\bottomrule

\end{tabular}
\end{table}

\subsection{Confidentiality and Access Constraints}
\label{sec:confidentiality}

The case study was carried out under a confidentiality agreement. The company requested full anonymity, including the non-disclosure of its name, precise geographical location, identifiable business units, and specific operational details that could enable indirect identification. To preserve confidentiality, contextual descriptions were deliberately reduced; only analytically relevant characteristics are retained.

The company permitted the use of anonymized interview and meeting quotations. Quotations were therefore edited for ensuring anonymity, without altering meaning. Identifiers such as proprietary system names, internal terminology, business unit names, or references to specific brands, products, or partners were removed or generalized.

\subsection{Data Analysis}
\label{sec:analysis}

We analyzed the interview transcripts inductively as soon as they became available, following Miles et al.~\cite{miles2014qualitative} guidelines. Accordingly, we conducted a First Cycle analysis, in which we selected ``chunks'' of data pertinent to our RQ and assigned them codes~\cite{miles2014qualitative}. Coding proceeded interview by interview, and codes were written in descriptive, participant-grounded language, staying close to participants' own wording; in-vivo codes were retained where participants' phrasing captured the phenomenon directly. The first author carried out the first iteration of analysis.

In the Second Cycle that followed, we synthesized the First Cycle codes into ``Pattern Codes''~\cite{miles2014qualitative}. In this integrative activity, we identified patterns across the earlier cycle codes by recognizing recurring themes that linked different codes together, whether through similarity or complementary contributions to a cohesive construct~\cite{miles2014qualitative}. Each First Cycle code was mapped
to the subtheme it evidences, and we documented this mapping, including the analytical decisions it required.

Through this process, the Pattern Codes progressively organized into the structure that frames our findings (see Fig.~\ref{fig:Model}); the sources of psychological costs, the organizational conditions surrounding the adoption, the psychological costs practitioners experience, and the responses through which they manage them (Sect.~\ref{sec:findings}). This structure was an analytical outcome of the Second Cycle rather than a framework imposed on it; the layers emerged as we examined how Pattern Codes related to one another, particularly the recurring distinction in the data between conditions that produce experiences and conditions that intensify them. Table~\ref{tbl:themes} documents examples of our analysis outcome.

\begin{table*}[ht!]
\footnotesize
    \caption{Examples of pattern codes, their sub-themes, and supporting evidence from the data}
    \label{tbl:themes}
    \renewcommand\arraystretch{1.0}

    \begin{tabular}{lp{3.2cm}p{8cm}}
    \toprule
      \textbf{Pattern codes} & \textbf{Sub-themes} & \textbf{Supporting evidence from the data}\\
      \hline

      \multirow{2}{*}{\shortstack[l]{\textbf{AI Disruption}}}
      & Fast Pace \& Emerging Technology & \emph{``we cannot catch up''} (P15).\\
      &  & \emph{``We dealing with an emerging technology which is moving very fast. So, we have to continue learning''} (P1).\\

      \hline

      \multirow{2}{*}{\shortstack[l]{\textbf{Organizational}\\ \textbf{AI Adoption}}}
      & Perceived adoption pressure & \emph{``every meeting is about AI ... I can't remember a company briefing or anything that was not about AI''} (P12).\\
      & Governance ambiguity & \emph{``it's[compliance obligation] clashing with what the organization want us to do''} (P19).\\

      \hline

      \multirow{3}{*}{\shortstack[l]{\textbf{Psychological}\\ \textbf{Costs}}}
      & Craft identity disruption & \emph{``I did not go to university to do this''} (P11).\\
      & Meaning and satisfaction erosion & \emph{``the scare for your job to become really boring, so that instead of building and creating stuff you are basically reviewing stuff, which nobody likes''} (P1).\\
      & Accountability anxiety & \emph{``if we weren't responsible for the code it produced, it would be a lot faster''} (P19).\\

      \hline

      \multirow{2}{*}{\textbf{Responses}}
      & Manage & \emph{``I code without AI often to maintain my skills''} (P3).\\
      & Mitigate & \emph{``we strip a lot of the data and ... feed it very little and use it like a more or less more advanced Google search''} (P12).\\

     \bottomrule

    \end{tabular}

\end{table*}

Once we finalized the list of Pattern Codes, we sought to identify the relationships among them. Through an iterative process of comparison, we revisited the coded data to examine how Pattern Codes interacted, co-occurred, and formed broader explanatory relationships. Consistent with Miles et al.'s conclusion drawing and verification activities, we iteratively examined alternative interpretations of the relationships among Pattern Codes and revisited the original interview excerpts to ensure that the emerging model remained empirically grounded~\cite{miles2014qualitative}.

This process revealed that the Pattern Codes did not represent independent themes but rather different components of a coherent process. Specifically, participants consistently described how characteristics of AI disruption and organizational AI adoption created conditions that gave rise to psychological costs, while simultaneously describing a range of responses through which they sought to manage, mitigate, or absorb these costs. We progressively refined these relationships into the conceptual model shown in Fig.~\ref{fig:Model}. The first author led the analysis effort, and the second author reviewed the coding and provided feedback.

\begin{table*}[t]
\footnotesize
\centering

\caption{Examples of empirical evidence supporting the relationships in the conceptual model.}
\label{tab:modelrelationships}
\renewcommand{\arraystretch}{1.0}

\begin{tabular}{p{6cm}p{1.5cm}p{6cm}}
\toprule
\textbf{Relationship} & \textbf{Verb} & \textbf{Representative empirical evidence} \\
\midrule

AI Disruption $\rightarrow$ Organizational AI Adoption
&
\textit{Shapes}
&
Participants described organizational AI adoption as reflecting the pace and discourse of the broader AI disruption. Examples included AI maturity targets mirroring industry narratives (P15), pervasive organizational AI messaging (P12), and governance struggling to keep pace with rapidly evolving technologies (P15, P19). \\

\addlinespace

AI Disruption $\rightarrow$ Psychological Costs
&
\textit{Gives rise to}
&
Participants reported psychological costs directly associated with AI itself, including identity disruption (P11), job-loss anxiety (P1), loss of authorship, reduced flow, and concerns over deteriorating understanding of their code (e.g., P6, P18 \& P19). \\

\addlinespace

Organizational AI Adoption $\rightarrow$ Psychological Costs
&
\textit{Amplifies}
&
Organizational AI adoption strategy intensified existing concerns through adoption pressure, productivity expectations, governance ambiguity, and conflicting organizational messages regarding AI use (P12, P19). \\

\addlinespace

Psychological Costs $\rightarrow$ Responses
&
\textit{Elicit}
&
Practitioners described adaptive responses tailored to specific costs, including oversight and verification practices to manage accountability (P19), deliberate manual coding to preserve skills (P3), identity reframing toward software delivery rather than coding (P10), and conservative AI use under governance uncertainty (P12). \\

\addlinespace

Accountability anxiety $\rightarrow$ Cognitive load intensification
&
\textit{Translates into}
&
To meet their accountability expectations, practitioners became invested in verification labor; reviewing AI output line by line because they remain answerable for it (P12), analyzing generated solutions before standing behind them (P2), and applying heightened scrutiny to AI-authored contributions in peer review (P16). The mechanism was stated directly; absent responsibility for the produced code, the work \emph{``would be a lot faster''} (P19). \\

\bottomrule

\addlinespace[2pt]
\multicolumn{3}{p{13.5cm}}{\scriptsize
\textit{Note.} The relational verbs reflect participants' accounts and the
patterns identified in the qualitative analysis; they are intended to preserve
the explanatory meaning of the empirical data. They do not imply causal
inference or causal effects and should be interpreted within the qualitative
design of the study.}
\\

\bottomrule
\end{tabular}
\end{table*}

\subsection{Unit of Analysis}
\label{sec:unit}

The unit of analysis in this study is practitioner's experience of the organizational
AI adoption, bounded within one organization. The adoption initiative provides the case boundary. This definition allowed us to examine how software practitioners experience AI adoption in SE practices and workflows within a specific organizational context.

This choice aligns directly with our research question, which seeks to understand the psychological costs practitioners experience and the practices through which they manage them. While the organizational AI adoption initiative constitutes the case boundary, it serves as the contextual setting rather than the object of analysis. Defining the unit of analysis at the practitioner level enabled us to examine how individuals interpreted, experienced, and responded to AI adoption while accounting for the shared organizational conditions within which these experiences were situated.

\subsection{Saturation}
\label{sec:saturation}

\noindent We sought meaning saturation, because our goal was explanatory understanding rather than simply cataloging codes~\cite{hennink2017code}. We monitored saturation throughout the analysis by iteratively comparing newly analyzed interviews with the evolving Pattern Codes and their meanings, and underlying supporting evidence~\cite{morse2004theoretical,aldiabat2018data}. We assessed whether additional interviews contributed substantively with new insights into the emerging sub-themes, Pattern Codes, or the relationships between them~\cite{hennink2017code}. 

As analysis progressed, subsequent interviews consistently reinforced and elaborated the existing conceptual structure without requiring new Pattern Codes or revisions to the emerging model (see Fig.~\ref{fig:Model}). At that point, we concluded that sufficient analytical depth had been achieved for the purposes of this study. Although all participants had been recruited before data collection commenced, we remained open to further recruitment had subsequent analysis revealed important conceptual gaps or insufficient development of the emerging findings.

Meaning saturation was not reached uniformly across the analytical categories. By the twelfth analyzed interview, all Pattern Codes except those concerning psychological costs were stable in their meanings and relationships. The psychological cost pattern gained conceptual depth and saturation at the eighteenth interview. Interviews analyzed after each respective saturation point (12 (all Pattern Codes) and 18 (Psychological Costs)) provided additional supporting evidence but did not alter the Pattern Codes or their relationships within the model.

\subsection{Member Checking}
\label{sec:member-checking}

\noindent We carried out member checking~\cite{birt2016member} interviews with participants. We emailed all informants who participated in the interviews, and twelve accepted to provide feedback (marked with (*) in Tbl.~\ref{tab:participants}). Prior to the interviews, participant received a 15-page report documenting the findings of this paper and other insights related to the organization's AI adoption strategy. The report served as pre-reading for the interviews. During the member checking interviews, we asked participants to comments on all reported findings.

Participants were asked to read the complete report before attending an individual member-checking interview. Using a semi-structured protocol, we discussed all findings in the report, including each psychological cost and the relationships represented in the conceptual model. For each finding, participants were asked whether it reflected their experience, whether any interpretation appeared inaccurate or overstated, and whether relevant experiences were missing. Feedback was provided orally during the interviews, audio recorded and transcribed. Participants generally regarded the findings relatable and consistent with their experiences. Their feedback did not require changes to our interpretations, new Pattern Codes or changes to the model's relationships, but it provided us with additional explanatory material. Table~\ref{tbl:checking} documents a few examples of the feedback we received in these interviews.

\begin{table*}[ht!]
\footnotesize
    \caption{Examples of comments from member checking interviews}
    \label{tbl:checking}
    \renewcommand\arraystretch{1.0}

    \begin{tabular}{lp{12cm}}
    \toprule
      \textbf{Participant} & \textbf{Feedback}\\
      \hline

      P6 & \emph{``I like how these psychological costs validate my experience'} (P6).\\
      
      P7&  \emph{``I agree largely with what has been said in the report''} (P7).\\

      P12 & \emph{``Some of your findings have been talked about in the corridors and during lunches, but we never openly confronted them''} (P12).\\

      P20 & \emph{``Some of the findings were expected, the report crystallizes them well''} (P20).\\

     \bottomrule

    \end{tabular}

\end{table*}

\section{Findings}
\label{sec:findings}

Our findings show that organizational AI adoption is experienced as more than the introduction of a new technology. Rather, participants experienced AI as a disruptive force that reshapes how they perceive their work, their professional identity, responsibilities, and their future within software engineering. Across the interviews, practitioners consistently articulated a set of psychological costs that emerged not only from adopting AI tools in their workflows, but also from the broader organizational AI adoption through which this technology is introduced and when AI is enacted in SE practice.

Participants described how characteristics of AI, including its rapid technological evolution, role-altering capability, and unknown risks, shapes how organizational AI adoption takes form, setting its tempo, its expectations, and its ambiguities. Both forces then converge on the practitioner, but not symmetrically. The disruption itself gives rise to a set of psychological costs, while the organizational adoption context amplifies them. These costs, in turn, elicited a range of responses through which
practitioners sought to manage, mitigate, or absorb their psychological impact. %

\begin{figure*}[!t]
    \includegraphics*[trim=3.5cm 0cm 4cm 2cm, clip, width=1.0\textwidth]{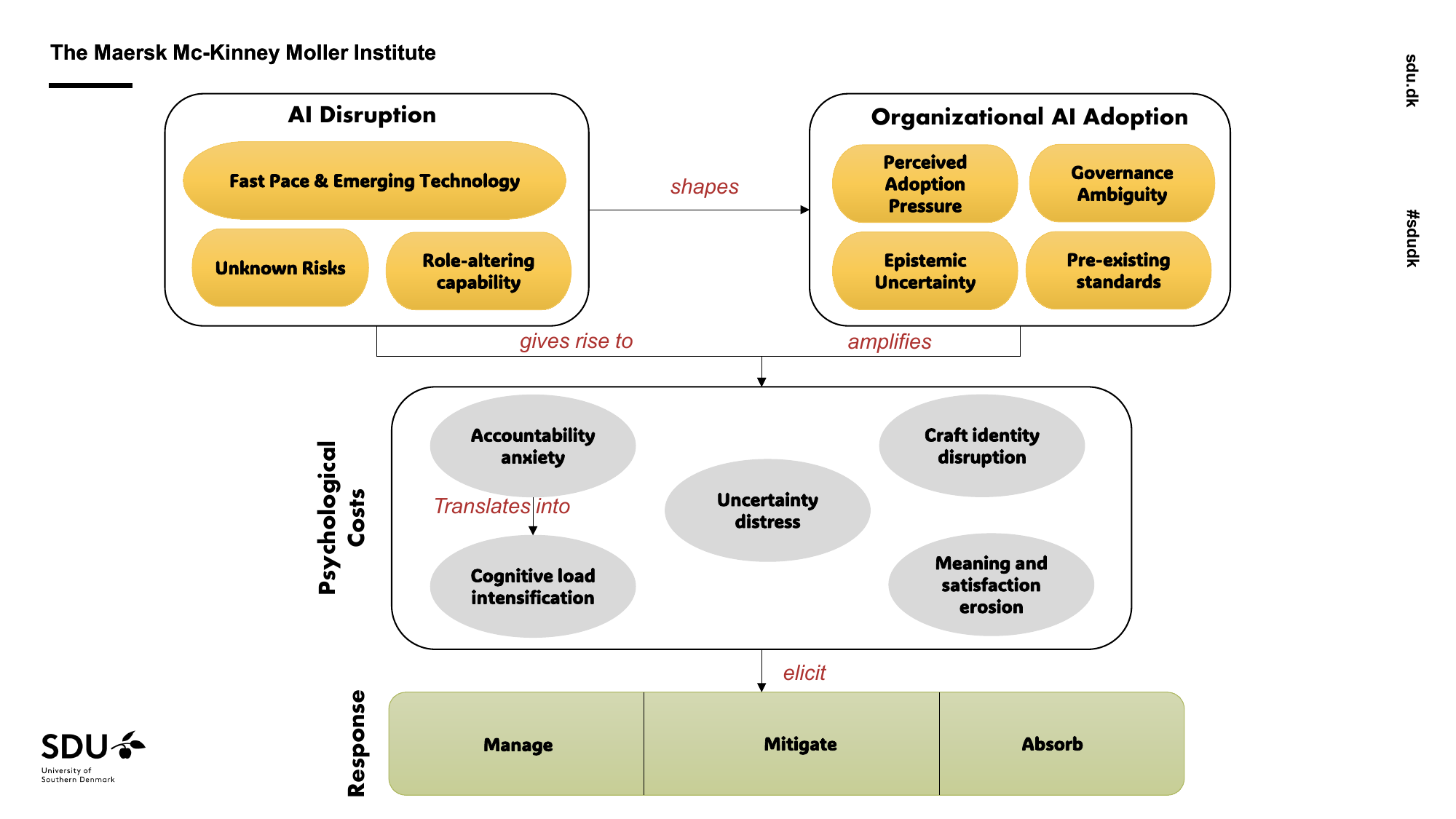}
        \caption{Conceptual Model of Psychological Costs of AI Adoption for Software Practitioners.}
        \label{fig:Model}
        
\end{figure*}

Figure~\ref{fig:Model} presents the conceptual model that emerged from the analysis and serves as the organizing framework for the remainder of this section. It illustrates how AI characteristics (e.g., evolutionary, unknown risks) give rise to distinct psychological costs and how practitioners respond through adaptive practices. The conceptual model in Figure~\ref{fig:Model} is presented as an integrated explanation of practitioners' experiences. To make the underlying mechanisms analytically transparent, the following sections examine each psychological cost individually, tracing how specific characteristics of AI disruption give rise to that cost, how organizational AI adoption amplifies it, and how practitioners respond. These analyses reconstruct the integrated model presented in Fig.~\ref{fig:Model}. 

\subsection{Uncertainty Distress}

\noindent Our participants described AI not as another software engineering tool, but as the beginning of a technological shift whose trajectory was difficult to comprehend. Unlike previous technological changes that could be evaluated against established engineering practices, AI was perceived as moving too quickly, carrying uncertain implications for software work, and accompanied by widespread societal narratives about its transformative potential, e.g., \emph{``we are working with technology that is moving just as fast as AI models are''} (P19).

Participants described struggling to keep pace with a technology that appeared to change faster than they could develop stable working practices. For example, new models' capabilities, and features emerged continuously, making it difficult to determine whether today's learning efforts would remain valuable tomorrow. P4 explained: \emph{``I need some more hours in trying to keep up with Claude and what is happening, because a lot of things is happening. New models are coming out, new versions of the model ... I try to use some of the free time I get to try to keep up with the AI technology, because it's really moving fast''} (P4). P2 highlighted the implications of this rapid technological evolution: \emph{``experiment and tell us what you think would be no longer a valid strategy now''} (P2); implying that the organizational adoption strategy of \emph{``experiment-and-share''} is not sustainable when the technology is a moving target.

This characteristic of AI may suggest that our participants experienced AI adoption in an environment where \emph{\textbf{knowledge becomes perishable}}. Unlike previous software technologies, where expertise could stabilize into established practices, AI continually invalidates emerging routines before they mature. Consequently, adaptation becomes an ongoing behavioural requirement rather than a temporary transition associated with learning a new tool. Psychologically, this left practitioners unable to form stable expectations about the value of their knowledge, the adequacy of their skills, and the trajectory of their professional development. As uncertainty became persistent rather than transient, it evolved from an informational challenge into a source of \emph{uncertainty distress}. P7 reflected on his worry: \emph{``It's a bit uncertain for me. I worry, how can I establish a consistent workflow in this volatility''} (P7). Consistent with our definition of psychological costs, uncertainty distress arose when practitioners appraised the persistent uncertainty surrounding AI adoption as exceeding their psychological resources to predict, prepare for, and adapt to future changes.

\begin{figure*}[!t]
    \centering
    \includegraphics[trim=8cm 6cm 2cm 4cm,clip,width=\textwidth]{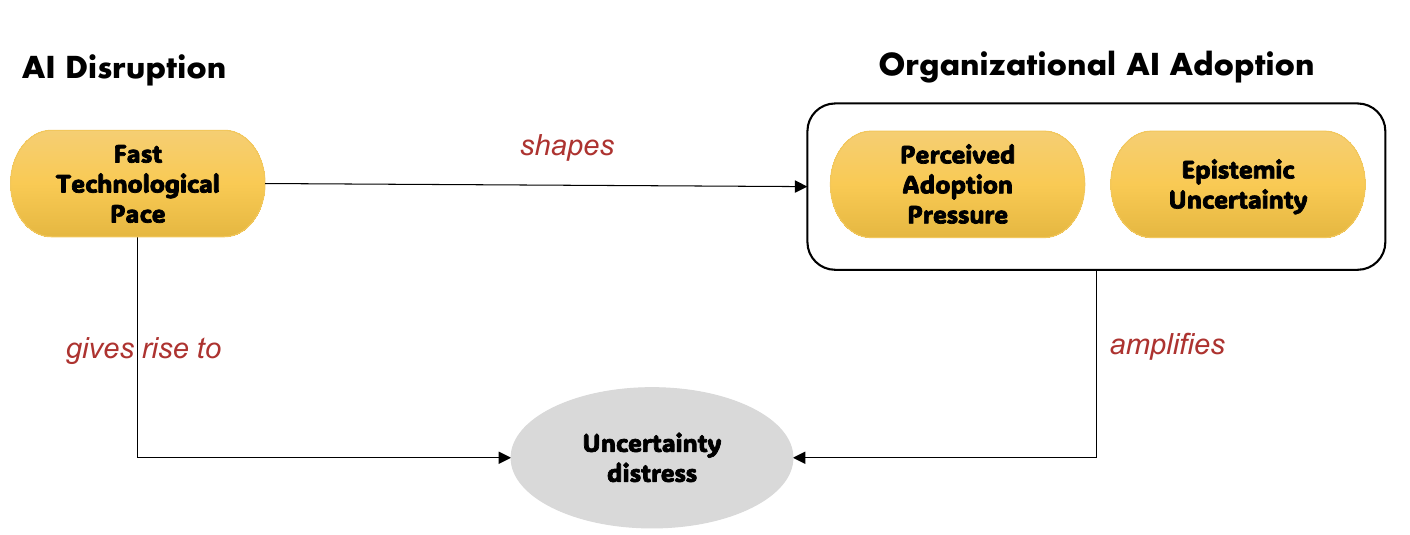}
    \caption{Conceptual Pathway Leading to Uncertainty Distress.}
    \label{fig:uncertainty_distress}
\end{figure*}

This pathway is illustrated in Fig.~\ref{fig:uncertainty_distress}. Uncertainty distress is given rise by the rapid pace of AI evolution and was further amplified by epistemic uncertainty and perceived adoption pressure within the organization. Nobody knew what best practices are and for some workflows like testing, practitioners struggled to identify proper use cases. When P9 was asked whether she benefited from the organizational strategy ``experiment-and-share,'' she reflected:  \emph{``the ones[experiment results] that were shared across a lot of us weren't really relevant for my role. I think, in general, it's hard to find the really good use cases in testing ... I don't think I found any use cases that, test wise, make my work more efficient''} (P11).

Practitioners in our sample seem to perceive AI adoption pressure as an amplifier of their uncertainty distress. While managers largely framed the AI maturity initiative as evidence of successful organizational adoption, practitioners experienced it as a source of uncertainty. P13 described how employees progressively advanced through the organization's AI maturity levels, observing that \emph{``a lot of them has moved to step 3 and 4. Some of them also to 5 running full agentic mode... I have one developer stating that I haven't wrote one single line of code the past month. So it's really getting under the skin on the developers''} (P13). Similarly, P15 characterized the initiative as encouraging practitioners to \emph{``climb the ladder of AI''} (P15), reflecting an organizational aspiration toward increasingly sophisticated AI use.

However, from practitioners' perspective, this emphasis on advancing AI maturity introduced uncertainty. P1 stated: \emph{``It's very offensive and uncomfortable now of course with all the measuring of people's AI maturity and that it's all over the news that this is causing people to be fired in a big way I'm sure everybody is also motivated to use AI by fear''} (P1). This assessment process seems to be perceived as a pressure mechanism; P19 explained: \emph{`` ... so they defined maturity levels, and up to a level 5, which is using agents to do basically what needs to be done. So they're pushing us to evaluate us. They are pushing us towards a certain level in the department. So they want people to be at level 4 where you are working with AI in a certain way''} (P19). Rather than reducing uncertainty, the maturity assessment introduced new uncertainties about what constituted successful AI adoption and whether failing to progress would carry professional consequences. In doing so, organizational expectations became another source of uncertainty beyond the technology itself.

In sum, the evolutionary and fast-pace nature of AI, coupled with the absence of stable organizational knowledge, and perceived adoption pressure, prevented practitioners from forming reliable expectations about how to invest in their competence and organize their work. This left them in a persistent state of anticipation and vigilance, continuously reassessing their preparedness for an uncertain future. The psychological cost therefore lies not in uncertainty itself, but in the sustained cognitive and emotional effort required to work with uncertainty that cannot be resolved. The ``cost'' materialized because practitioners expended their psychological resources trying to anticipate change, protect their professional competence, and make decisions without stable reference points. Consequently, uncertainty ceased to be a temporary feature of technological change and became an enduring source of psychological strain during organizational AI adoption.

\subsection{Accountability Anxiety}

\begin{figure*}[th!]
    \centering
    \includegraphics[trim=5cm 5cm 4cm 3cm,clip,width=\textwidth]{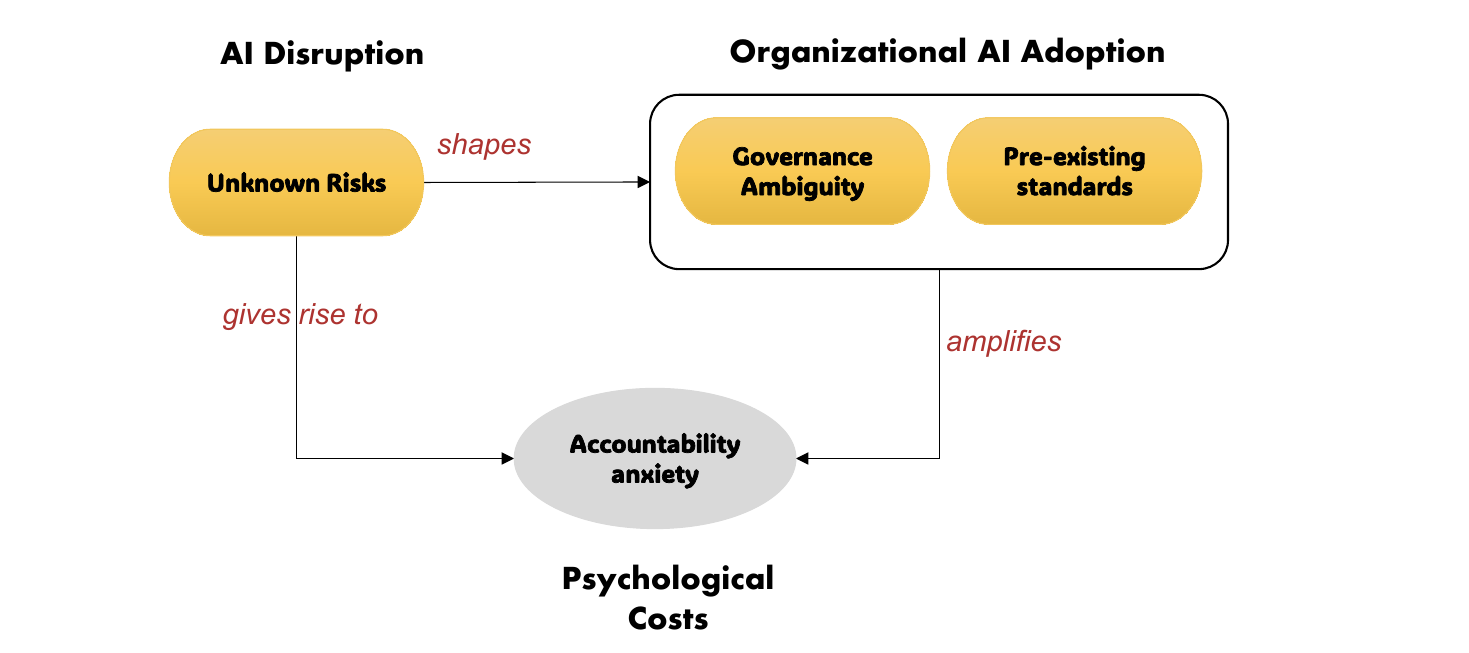}
    \caption{Conceptual Pathway Leading to Accountability Anxiety.}
    \label{fig:accountability_anxiety}
\end{figure*}

AI seems to introduce accountability anxiety to our participants, e.g., \emph{``not being able to fully vouch for the result''} (P7). This psychological effect appears to stem from the participants' professional confidence to take responsibility for the AI outputs. P2 explained: \emph{``I have a task right now, where I have to make our, uh, the Kotlin that shares the Android native and the iOS native ... it needs to receive notifications. And that needs to hook into the native code on the platform. I have absolutely no idea how to do that other than I know I need to connect to something on Android and something on iOS. But it has removed the stress of knowing that I have to do that tomorrow. Because I'll make AI analyze, I mean, Claude, make a deep analysis on how to do it and propose solutions for it. The problem with that is, and that gives me the anxiety of, is that even working? Can I stand behind it?''} (P2).

This accountability anxiety seems to originate from the unknown risks associated with integrating AI-generated outputs into production systems. P8 explained: \emph{``If you get an AI to develop a feature, and you don't know how it took the requirements and broke it down to stories, and you cannot tell the audit team how that happened ... I think there are some risks laying around''} (P8). P2 further explained this dual paradox, AI reduces his \emph{``load''} while introducing accountability anxiety: \emph{``It eases the load, but it increases the anxiety of doing something wrong''} (P2).

Accountability is also systemic at \textsc{SoftHouse}, where it is embedded in a culture of trust and professional responsibility. The organizational culture promotes a shared sense of accountability, i.e. \emph{``we trust you''} (P9). Rather than reducing responsibility, this trust places practitioners in the role of organizational gatekeepers, responsible for ensuring the quality and integrity of software before deployment. As a result, employees internalize organizational trust as a personal obligation to exercise sound professional judgment, making the trust relationship bidirectional. P6 explained: \emph{``I think it's because at SoftHouse, we have this, we trust you until everything else is proven wrong. So they put trust on us. They trust us to do our work and trust us to be the gatekeepers, you could say''} (P6). This trust-by-default stance creates a paradox: while AI reduces technical workload, it simultaneously increases the burden of personal judgment, as practitioners must now verify opaque AI outputs without clear organizational guardrails. In the absence of clear rules, the organization's trust-by-default stance converts every practitioner into their own compliance officer; it transfers the liability judgment onto the individual.

One year into the AI adoption, the organization is still lacking clear governance guidelines for AI use; P19 stated: \emph{``if they[SoftHouse] want to push the AI to the level of the maturity level that they want or envision us to be at, they need to provide us with the compliance guardrails that makes it possible''} (P19). Such ambiguity appears to amplify practitioners' anxiety for their accountability. P11 summarized this situation: \emph{``compliance concerns have been holding back people, what actually was allowed ... they could use AI, but they were uncertain if they actually did something that could pose a security or other compliance concern ... I have heard people say: I don't know what I'm allowed to do, so I don't do anything, to not run the risk that I do something I was not allowed to do''} (P11).

\textsc{SoftHouse} has high affinity for high standards, quality, and adherence to compliance requirements. P1 stated: \emph{``SoftHouse is a very process-strong house. And you can say that the commitment to delivering quality is very, very strong with us''} (P1). When P18 was asked about the objections he hears in his team, he replied: \emph{``people have been mostly concerned about quality they have been concerned to be pushed to use tools which did not provide quality since people really are committed to delivering quality ...  they are scared to be pushed to produce poor quality''} (P18). This evidence may suggest that quality at \textsc{SoftHouse} is an organizational objective translated into a professional norm that practitioners appear to internalize as part of their identity and responsibility. Consequently, their concerns were that relying on AI could compromise the quality standards for which they are accountable.

Figure~\ref{fig:accountability_anxiety} illustrates this pathway. Unknown risks associated with AI-generated outputs gave rise to accountability anxiety, while both governance ambiguity and \textsc{SoftHouse}'s pre-existing quality and compliance standards amplified this psychological cost. Because practitioners remained expected to satisfy these established standards despite relying on AI, they experienced heightened responsibility to verify, justify, and ultimately stand behind AI-assisted decisions.

AI seemed to constrain practitioners' ability to determine, understand, and justify how solutions were produced, while leaving their professional accountability unchanged. The psychological cost therefore lies in the depletion of agency resources (i.e., being in control of their work) required to exercise professional responsibility. Practitioners were expected to stand behind AI-assisted decisions despite experiencing diminished control over the reasoning, provenance, and rationale underlying those decisions.

\subsection{Cognitive load intensification}

\begin{figure*}[th!]
    \centering
    \includegraphics[trim=6cm 5cm 4cm 3cm,clip,width=\textwidth]{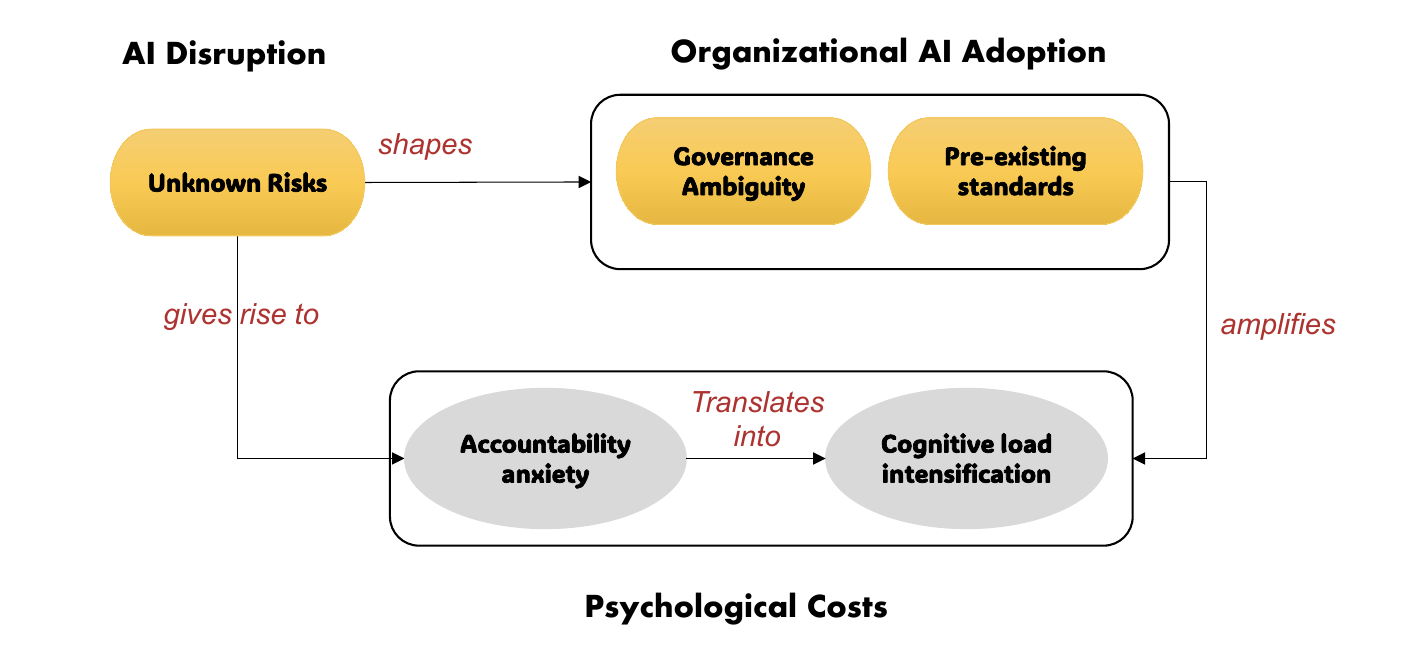}
    \caption{Conceptual Pathway Leading to Cognitive Load Intensification.}
    \label{fig:cognitive_load}
\end{figure*}

To meet their accountability expectations, our participants reported expending additional resources verifying, questioning, and justifying AI-generated outputs to fulfill their professional obligations. P12 explained: \emph{``I have to do more work. And it hasn't eased my cognitive load, actually. It's made me ... I have to work harder, I feel like, with AI. Because now you can do more, so you feel like you need to do more ... I need to look through the result line by line, I'm accountable for it''} (P12).

Here, cognitive load intensification arose when practitioners appraised AI-assisted work as requiring greater cognitive effort than their available mental resources could comfortably sustain (i.e., \emph{``work harder''}). AI shifted practitioners' effort from producing code to verifying, evaluating, and justifying AI-generated outputs to meet their accountability obligations. As illustrated in Fig.~\ref{fig:cognitive_load}, accountability anxiety translated into cognitive load intensification by requiring practitioners to continually verify, question, and justify AI-generated outputs before they could confidently stand behind the work they produced in collaboration with AI.

\subsection{Craft Identity Disruption}

\begin{figure*}[th!]
    \centering
    \includegraphics[trim=3.5cm 8.5cm 4cm 6cm,clip,width=\textwidth]{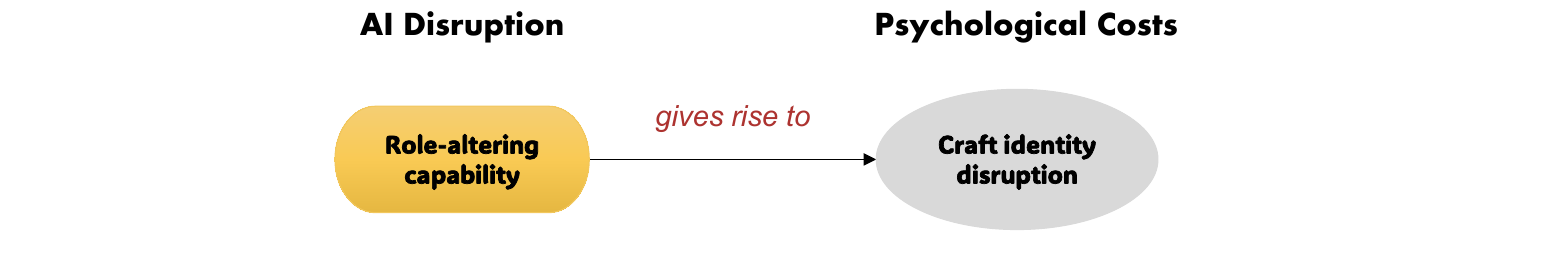}
    \caption{Conceptual Pathway Leading to Craft Identity Disruption.}
    \label{fig:craft_identity}
\end{figure*}

The accountability question touches on something deeper than quality checking. P2 is asking whether AI-assisted work is authentically his own in a professional sense: \emph{``Can I stand behind it once it has made it?''} (P2). His description of his own practice carries the same tension: \emph{``I asked AI and then I analyzed it and it seems like a good way''} (P2); the ownership he can claim is evaluative rather than generative (i.e., ``I did not build it, I evaluated it''); he endorses the solution, but did not author it.

As illustrated in Fig.~\ref{fig:craft_identity}, identity seems to be disrupted by AI performing the core activities that defined our practitioners' professional selves. This role-altering capability was echoed across almost all participants, e.g. P2 stated: \emph{``It's very powerful, very powerful. I can make big, big, big tasks with it''} (P2) and \emph{``AI is relieving my workload by summarizing large files faster than humanly possible''} (P12). This shift disrupted the meaning underlying the identity of software practitioners in our sample; they are questioning what their expertise now amounted to: \emph{``then you just have to prompt ... I did not go to university to do this''} (P11).

This role-altering capability fundamentally challenged how practitioners understood their professional identity; this was more prominent among software engineers, testers, and architects to varying degrees (see Sect.\ref{sec:roles}). For many participants, software engineering was not simply an occupation but an important source of self-definition, where producing code and other software artifacts represented both professional competence and personal accomplishment. As AI increasingly assumed the generative aspects of software development, practitioners described a growing disconnect between the work they performed and the work that had previously defined them. P7 (a software engineer) reflected on this shift: \emph{``I find it difficult to separate myself as a person from the work that I do... being a software engineer who writes code is sort of who I was, and now I'm a software engineer who prompts an AI and then reads the code''} (P7). P17 (a software architect) sees a drastic shift in his role, \emph{``I'm a technical guy ... it[AI]'s taking away all the fun parts ... And now I'm just a prompt monkey ... I really love the fact that I can actually still revisit and validate at code level. I don't have that luxury when I work with AI tools''} (P17).

Rather than merely changing tasks, AI altered the basis upon which practitioners experienced authorship, craftsmanship, and professional pride. P12 further described how \emph{``the final product feels like something I didn't make anymore because it's written by some machine''} (P12). In this sense, practitioners no longer questioned only the value of their technical expertise; they questioned whether the work they now performed remained authentic to the professional identity they had spent years developing.

For testers, the disruption took a different form: rather than losing generative work, they experienced the worth and standing of their role being renegotiated. P10, an automation test engineer, pushed back against efficiency as the measure of AI's value in her role, \emph{``mostly I think the purpose should be to strengthen the quality of tests''} (P10), repositioning the tester as the human checkpoint whose importance grows as developers delegate more of their work upstream. P1, a test manager, was blunter about the role's standing during the transition: \emph{``a tester is easier
to fire than an architect''} (P1); she declined to experiment with more advanced AI integrations at work, protecting her professional standing by judging that the consequences of a mistake would fall harder on her role than on others.

This disruption extended beyond the present into practitioners' anticipated professional future. P7 worried that prolonged reliance on AI would erode the very skills that once distinguished him professionally, wondering \emph{``if we lose access to these things ... will I still be good at it?''} (P7). Similarly, P1 anticipated a future in which software engineering would shift from creating software to supervising AI-generated work, describing concerns that the profession would become \emph{``really boring''} (P1) and expressing fears of \emph{``skill atrophy''} (P1) that might ultimately leave practitioners unable even to critically review AI-generated solutions. 

Our analysis suggests that craft identity disruption reflects the erosion of a coherent professional self rather than simply changes in job tasks. As AI assumed activities that practitioners regarded as the essence of software engineering, they experienced increasing difficulty maintaining continuity between who they had become through years of practice and who their work now required them to be. The psychological cost therefore lies in the sustained effort required to reconstruct a professional identity whose traditional foundations, craftsmanship, authorship, and technical mastery, have become increasingly constrained. Consistent with our definition of psychological costs, practitioners appraised these changes as threatening valued aspects of their professional identity while exceeding their psychological resources to preserve a coherent sense of who they are as software professionals.

\subsection{Meaning and Satisfaction Erosion}

\begin{figure*}[th!]
    \centering
    \includegraphics[trim=3.5cm 8.5cm 4cm 6cm,clip,width=\textwidth]{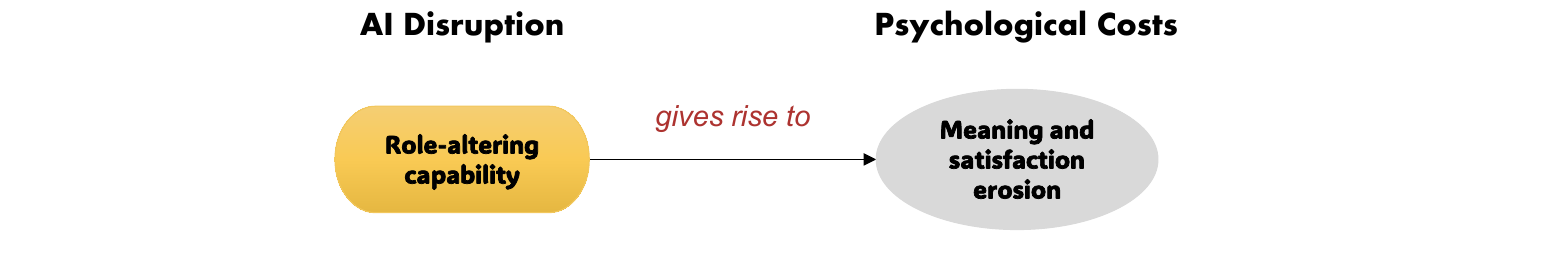}
    \caption{Conceptual Pathway Leading to Meaning and Satisfaction Erosion.}
    \label{fig:meaning}
\end{figure*}

As illustrated in Fig.~\ref{fig:meaning}, the role-altering capability of AI appears to challenge our participants' professional identity, thereby reshaping the meaning and satisfaction practitioners derive from their work. P16 vividly explained: \emph{``I felt a certain pride at being good at coding myself at one point, and now it feels a little bit like the mechanical parts that I enjoyed of using shortcuts well, navigating the code base well. All of these things that I worked on quite hard are not that important anymore''} (P16). P7 echoed a similar sentiment: \emph{``I always thought the real part of the work was producing the code. That's what we deliver, and that's what the customer wants, that's what they use, and that's how we identify ourselves. If something doesn't work, it's because there's something wrong in the code usually ... I think I just made it a very intimate part of myself, and being a software engineer who writes code is sort of who I was, and now I'm a software engineer who prompts an AI and then reads the code''} (P7). This sentiment was felt among almost all software engineers in our sample, e.g., \emph{``... the final product feels like something I didn't make anymore. and that's a weird experience sometimes when you are used to owning the entire process and now the actual thing that you deliver, isn't written by me anymore. It's written by some machine that understood maybe part of what was in my head at some point.''} (P14).

Our analysis suggests that meaning and satisfaction erosion seems to reflect more than a change in work activities; it reflects a change in how software engineers experience the intrinsic value of their work. Activities that previously provided a sense of craftsmanship~\cite{alami2024understanding,alami2025accountability}, accomplishment, and personal fulfillment became increasingly delegated to AI, leaving practitioners with a diminished sense of ownership over the outcomes they delivered. Rather than deriving satisfaction from creating software, software engineers in our sample appear to increasingly describe evaluating, reviewing, and approving software created by AI. The psychological cost therefore lies in the gradual erosion of work as a source of meaning, accomplishment, and professional fulfillment. Aligned with our definition of psychological cost, software engineers appraised this transformation as diminishing the motivational and emotional rewards they associated with developing software, requiring them to reconcile a profession they once found deeply meaningful with one that seems no longer to provide the same sense of purpose and satisfaction.

This erosion was not uniform across roles; it concentrated where the displaced activity had been the role's craft (i.e., software code and other artifacts part of the software product). Testers, whose craft is evaluative rather than generative, described the shift toward verification as an intensification of their role rather than its hollowing; P10 reported unchanged enjoyment of AI-assisted
work, \emph{``When make the test strategy for a feature. It[AI] ensures that we actually get to the stuff we wanted to test. So when I do a test implementation, AI helps me identify the gap in my tests, it complements my effort ... And I really like working with it. I don't have any resistance against it. I think it's nice''} (P10).

Architects presented the mirror case; they were among the most enthusiastic adopters, as AI extended their generative reach: \emph{``I suddenly had a developer in my back pocket''} (P17); referring to Claude Code being the ``developer'' becoming available to assist in prototype development. An enthusiasm other participants attributed to architects prototyping without developer pushback, \emph{``they don't have a developer
saying no to them''} (P12).

\medskip
\medskip

\begin{mdframed}[
  leftline=true,
  rightline=false,
  topline=false,
  bottomline=false,
  linewidth=3pt,
  linecolor=black,
  innerleftmargin=10pt,
  innertopmargin=6pt,
  innerbottommargin=6pt
]

\noindent\textbf{Psychological Costs of Organizational AI Adoption in SE.} In response to our RQ (i.e., psychological costs), practitioners experienced five psychological costs: \emph{uncertainty distress}, \emph{accountability anxiety}, \emph{cognitive load intensification}, \emph{craft identity disruption}, and \emph{meaning and satisfaction erosion}. These costs may reflect the depletion of psychological resources required to adapt to organizational AI adoption while continuing to enact their professional role. Our findings suggest that organizational AI adoption is not experienced solely as a technological or organizational change, but also requires adaptation at the human level, as practitioners renegotiate their professional responsibilities, identity, the meaning of their work, and their relationship with AI.

\end{mdframed}

\subsection{Role-based Analysis}\label{sec:roles}

Our analysis shows that the psychological costs are uneven across the roles we covered in our data collection; while some costs are felt evenly, others are rather role-specific. To examine how psychological costs vary across roles, we complemented our thematic analysis with a role-based magnitude assessment~\cite{miles2014qualitative}. For each psychological cost, we determined the number of participants within each role whose interviews contained at least one First Cycle code mapped to that cost. We then expressed this as the prevalence, how widespread a cost is within a role in our sample. 

Table~\ref{tab:role-prevalence} documents this analysis. Three properties of this system should be noted. First, it counts participants, not coded segments; a participant who voiced a cost repeatedly and one who voiced it once contribute equally, which guards against differences in interview length and participant expressiveness. Second, the qualitative prevalence in this sample supports no statistical comparison across roles; we report the underlying fractions so readers can assess the evidence directly. 

In our sample, software engineers have the highest fractions for four psychological costs. One possible interpretation is that the psychological impact of AI adoption is associated with the extent to which a role depends on directly creating software artifacts as the primary expression of professional expertise. Software engineers enact their professional agency through writing, shaping, and owning code. As AI increasingly performs these activities, engineers experience greater disruption to the mechanisms through which they exercise competence, derive meaning, and demonstrate professional value. In contrast, roles whose primary responsibilities involve coordination, architectural decision-making, governance, or organizational oversight appear less directly affected because AI augments rather than substitutes the activities through which these practitioners enact their professional role.

\begin{table*}
\caption{Prevalence of psychological costs by role. Prevalence is the share of a role's participants whose interviews contained at least one First Cycle code. Codes expressing costs only as reported speech about others were not counted toward the reporting participant.}

\label{tab:role-prevalence}
\footnotesize

\begin{tabular}{llc}
\toprule
\textbf{Role} & \textbf{Psychological cost} & \textbf{Prevalence} \\
\midrule

\multirow{6}{*}{\shortstack[l]{Senior Management\\(n=5)}}
 & Accountability anxiety & 2/5  \\
 & Craft identity disruption & 0/5  \\
 & Meaning and satisfaction erosion & 0/5  \\
 & Cognitive load intensification & 0/5  \\
 & Uncertainty distress & 3/5  \\
\midrule

\multirow{6}{*}{\shortstack[l]{Software Architecture\\(n=3)}}
 & Accountability anxiety & 0/3  \\
 & Craft identity disruption & 1/3  \\
 & Meaning and satisfaction erosion & 1/3  \\
 & Cognitive load intensification & 1/3  \\
 & Uncertainty distress & 0/3  \\
\midrule

\multirow{6}{*}{\shortstack[l]{Testing\\(n=3)}}
 & Accountability anxiety & 1/3  \\
 & Craft identity disruption & 0/3  \\
 & Meaning and satisfaction erosion & 1/3  \\
 & Cognitive load intensification & 0/3  \\
 & Uncertainty distress & 3/3  \\
\midrule

\multirow{6}{*}{\shortstack[l]{Software Engineers\\(n=10)}}
 & Accountability anxiety & 10/10  \\
 & Craft identity disruption & 10/10  \\
 & Meaning and satisfaction erosion & 10/10  \\
 & Cognitive load intensification & 7/10 \\
 & Uncertainty distress & 10/10 \\

 \toprule

\end{tabular}
\end{table*}

We interpret these role-based patterns as suggesting that psychological costs may be more pronounced when AI displaces activities central to practitioners' professional agency. In other words, it could be that AI use alone may not predict psychological costs; but also how much AI replaces the activities through which practitioners exercise their professional agency. However, for testers and architects, such patterns do not materialize to the same extent. For both roles, the representation in our sample is low. In addition, for testing AI support is complementary, yet, uncertainty distress is notably high. This aligns with our earlier finding that testers struggle to identify stable, efficient use cases for AI in their workflows, leaving them in a state of persistent epistemic uncertainty despite the technology being complementary rather than substitutive. For architects, the agency displacement could be negligible, as architectural decision-making remains a human task, at least in the context of \textsc{SoftHouse}. Finally, for senior management, accountability anxiety and uncertainty distress are highly prevalent. This can be explained by their direct ownership of the AI adoption initiative and the organizational pressure to demonstrate successful implementation.

\medskip

\begin{mdframed}[
  leftline=true,
  rightline=false,
  topline=false,
  bottomline=false,
  linewidth=3pt,
  linecolor=black,
  innerleftmargin=10pt,
  innertopmargin=6pt,
  innerbottommargin=6pt
]

\noindent\textbf{Role-Based Variation in Psychological Costs.} Psychological costs were not experienced uniformly across the roles in the scope of our study. Our findings suggest that their prevalence is associated with \textbf{agency displacement}, which we define as the extent to which AI assumes the activities through which practitioners traditionally exercise professional judgment, expertise, and ownership. Among \textbf{Software Engineers}, whose professional agency is strongly enacted through designing and producing software, AI appeared to displace a substantial portion of these activities and was associated with the broadest range of psychological costs. By contrast, for \textbf{Software Architects}, AI primarily augmented rather than substituted their core responsibilities, corresponding to comparatively limited experiences of craft identity disruption and meaning erosion. For \textbf{Testers}, AI also complemented rather than displaced their work, yet the absence of stable and role-relevant AI use cases was associated with pronounced uncertainty distress. Finally, \textbf{Senior Management} experienced elevated accountability anxiety and uncertainty distress, reflecting their responsibility for leading organizational AI adoption while navigating its uncertain trajectory.

\end{mdframed}

\subsection{Responses to Psychological Costs}

We also observed that participants respond to these costs in different ways of dealing with them in their everyday practices. Although participants described a wide variety of responses, these did not differ merely in their form but in the function they served. We therefore organized them according to how practitioners related to the psychological costs they experienced rather than according to the specific behaviors themselves. \emph{\textbf{Managed}} responses comprised practices that enabled practitioners to continue working alongside AI while keeping psychological costs under control. \emph{\textbf{Mitigated}} responses reduced practitioners' exposure to particular psychological costs or limited their potential impact. In contrast, \emph{\textbf{absorbed}} responses reflected situations in which practitioners perceived the psychological cost as unavoidable and continued to work while consciously living with its consequences. This functional organization highlights that practitioners did not simply react to AI adoption; they responded differently depending on whether they believed a psychological cost could be controlled, reduced, or had to be endured. These responses illustrate that practitioners' experiences of AI adoption include ongoing adjustments in how they define, negotiate, and enact the boundaries of AI-human collaboration.

Our analysis shows that these responses are efforts to preserve valued psychological resources: \emph{agency} \textbf{(being in control)}, \emph{competence} \textbf{(being capable)}, and \emph{professional identity} \textbf{(being the kind of engineer they value being)}. Table~\ref{tab:resources-responses} documents the psychological resources being protected by the responses and their corresponding costs.

\paragraph*{Agency} Practitioners seem to attempt preserving their agency by managing the extent to which AI could influence their work while retaining themselves as the decision maker. Rather than accepting AI outputs at face value, they described deliberately being in charge of the decision process through verification, review, and staged delegation. P2 explained that every AI-generated solution was followed by an ``analysis'' before adoption: \emph{``I analyzed it and it seems like a good way''} (P2). Similarly, P12 explained, \emph{``I need to look through the result line by line, I'm accountable for it''}. When uncertainty exceeded acceptable levels, practitioners shifted from managing to mitigating the cost by limiting AI's involvement altogether. P9, for example, rejected direct integrations outright, stating \emph{``MCP integration is so far a no-go. I won't do that, because I won't get fired''} (P9), while P12 removed sensitive data before using cloud-based AI tools. These accounts show that preserving agency did not mean rejecting AI, but regulating its role so that practitioners retained sufficient control to exercise professional judgment.

\paragraph*{Competence} Responses aimed at preserving competence reflected practitioners' concern that prolonged dependence on AI could gradually erode their own technical capability. Rather than allowing AI to replace the learning process, participants deliberately created opportunities to continue exercising their skills. P7 described beginning each day with a period of manual coding because \emph{``if you don't use it, you lose it''} (P7), treating coding itself as a practice that required continual maintenance. Similarly, P10 intentionally attempted solutions before consulting AI to preserve independent reasoning, while P20 deliberately limited delegation because \emph{``if you use it too much, you will forget some of the skills''} (P20). P18 went even further by dedicating specific time in his coding effort to \emph{`` ... manual coding without AI''} (P18). These responses suggest that practitioners viewed competence as something that required deliberate preservation through continued engagement with core engineering activities. This could also be an investment in maintaining agency through preserving skills required to validate AI outputs.

\begin{landscape}
\begin{table}

\footnotesize
\centering

\caption{Psychological resources preserved through practitioners' response patterns, the costs addressed, the responses enacted, and illustrative evidence.}
\vspace{-0.3cm}

\label{tab:resources-responses}
\renewcommand{\arraystretch}{0.90}
\footnotesize

\begin{tabular}{p{5cm}p{2.5cm}p{5cm}p{6.2cm}}
\toprule

\textbf{Psychological Resource Preserved} & \textbf{Response Pattern} & \textbf{Psychological Cost(s) Addressed} & \textbf{Responses} \\
\midrule

\multirow{2}{=}{\textbf{Agency}\newline (being in control)}
& \textit{Manage}
& Accountability anxiety \newline
  Cognitive load intensification
& Oversight and comprehension-preserving practices: \newline
  1. staying in the loop; \newline
  2. prior codebase knowledge as oversight prerequisite; \newline
  3. complexity-based delegation triage; \newline
  4. plan and diff review; \newline
  5. baseline testing before accepting changes; \newline
  6. deliberate deceleration; \newline
  7. code-level verification (e.g., ``line-by-line verification'')
\\
\cmidrule(lr){2-4}

& \textit{Mitigate}
& Uncertainty distress %
& Risk containment and exposure limitation: \newline
  1. self-imposed limits pending explicit permission; \newline
  2. discontinuation of unsanctioned experiments; \newline
  3. data stripping before cloud tool use; \newline
  4. partitioning work into safe and gray zones; \newline
  5. declining write-capable integrations
 \\
\midrule

\textbf{Competence}\newline (being capable)
& \textit{Mitigate}
& Uncertainty distress %
& Skill preservation practices: \newline
  1. deliberate periodic manual coding ritual; \newline
  2. self-then-AI sequencing (own draft before AI to protect learning); \newline
  3. AI delegation moderation (boilerplate only, retaining business logic heavy work); 
 \\
\midrule

\multirow{2}{=}{\textbf{Professional Identity}\newline (being the kind of engineer they value being)}
& \textit{Mitigate}

& Craft identity disruption
& Identity reframing: \newline
  1. role redefinition toward orchestration; \newline
  2. relocating professional worth from writing code to delivering software; \newline
  3. positioning the human as guardian of architectural integrity; \newline
  4. reframing SE as problem solving rather than coding
\\
\cmidrule(lr){2-4}

& \textit{Absorb}
& Craft identity disruption \newline
  Meaning and satisfaction erosion \newline
  Cognitive load intensification
& Resigned adaptation and compensation: \newline
  1. loss of enjoyment endured; \newline
  2. enjoyment displaced outside paid work; \newline
  3. continuing to work within the new conditions without resolution; \newline
  4. capitulation to AI adoption
\\
  
\bottomrule

\end{tabular}

\vspace{4pt}
\raggedright
\footnotesize 

\textbf{Methodological note:} We employed a multi-level functional analysis to conclude this mapping. First, First Cycle coding identified granular practices (e.g., line-by-line verification). Second, we aggregated these practices according to their intended function (i.e., manage, mitigate, and absorb). Third, we traced each functional pattern to the psychological resource it aimed to preserve (agency, competence, or professional identity). Finally, we linked each response pattern to the specific psychological cost(s) it addressed by examining which resource was threatened and whether the response's function directly countered that cost's mechanism. \\

\end{table}
\end{landscape}

\paragraph*{Professional Identity} Preserving professional identity proved considerably more complex because, unlike agency or competence, identity could not always be restored through changes in work practices. Some practitioners mitigated identity disruption by redefining what it meant to be a software engineer. P10 distinguished between \emph{``delivering software, not delivering code''} (P10), relocating professional worth from code production to solving customer problems. Others repositioned themselves as orchestrators of AI or guardians of architectural integrity, with P19 describing the uniquely human contribution as \emph{``seeing the missing parts in a solution or a setup''} (P19). However, several participants perceived the loss as irreversible and instead absorbed its psychological consequences. P19 acknowledged that losing the enjoyment of coding was \emph{``not a nice feeling''} (P19). Likewise, when asked how he dealt with the increased mental burden of AI-assisted work, P7 replied simply, \emph{``I don't. I just keep going''} (P7). These accounts may suggest that when practitioners could no longer preserve valued aspects of their professional identity, they continued working by accepting the psychological costs as part of the evolving nature of software engineering rather than expecting them to disappear.

These responses may also suggest that practitioners sought to preserve different psychological resources depending on the nature of the psychological cost they experienced. \emph{\textbf{Agency}} was primarily preserved by changing how work was performed, enabling practitioners to retain control, oversight, and professional judgment while collaborating with AI. \emph{\textbf{Competence}} was preserved by changing how learning occurred, deliberately maintaining opportunities to exercise and maintain technical expertise independently of AI. \emph{\textbf{Professional identity}}, however, proved less amenable to behavioral adjustment. Practitioners appear to either reconstruct what it meant to be a software engineer or practitioner in AI-assisted software engineering or, when such reconstruction was not possible, continued working while accepting enduring losses to craftsmanship, meaning, and satisfaction. 

\medskip

\begin{mdframed}[
  leftline=true,
  rightline=false,
  topline=false,
  bottomline=false,
  linewidth=3pt,
  linecolor=black,
  innerleftmargin=10pt,
  innertopmargin=6pt,
  innerbottommargin=6pt
]

\noindent\textbf{Managing Psychological Costs.} Practitioners responded to psychological costs by preserving three core psychological resources: \emph{agency} (being in control), \emph{competence} (being capable), and \emph{professional identity} (being the kind of engineer they value being). These resources, however, required different forms of response. To preserve \textbf{agency}, practitioners primarily \emph{managed} costs by regulating AI's role through practices such as staged delegation, deliberate oversight, and line-by-line verification, enabling them to remain in control. To preserve \textbf{competence}, they \emph{mitigated} psychological costs by deliberately maintaining opportunities to exercise their own skills, for example through manual coding rituals and limiting AI-generated first drafts. \textbf{Professional identity}, however, proved less amenable to behavioral adjustment. Practitioners either \emph{mitigated} identity disruption by redefining what it meant to be a software engineer (e.g., from producing code to orchestrating software development) or \emph{absorbed} enduring losses to craftsmanship, meaning, and enjoyment when such reconstruction was no longer possible.

\end{mdframed}

\section{Discussion and Implications}
\label{sec:discussion}

\noindent We commence this section by discussing our contribution, then the theoretical and practical implications of our findings. 

\subsection{Contribution}

Our study contributes an explanatory and diagnostic model of psychological costs during organizational AI adoption in SE. Our findings overlap with established technostress concerns, including uncertainty, overload, threatened competence, and reduced satisfaction~\cite{ayyagari2011technostress,ragu2008consequences}. Recent SE research also recognizes the verification demands and continued human responsibility accompanying GenAI use~\cite{jeyam2026still}. \textbf{Our contribution lies in explaining how these demands become psychological costs through their interaction with engineering obligations, professional identity, and organizational adoption conditions.} Table~\ref{tab:contribution} positions the identified psychological
costs in relation to their closest technostress constructs and specifies
the explanatory contribution of our model. 

Our model (Fig.~\ref{fig:Model}) distinguishes characteristics of AI, organizational conditions that give raise and intensify the resulting costs, and responses through which practitioners attempt to preserve agency, competence, and professional identity. These relationships explain how AI adoption inherent conditions translate into human burdens. For example, the model distinguishes accountability pathway from changes to professional identity and meaning. When AI assumes activities through which practitioners previously experienced authorship, mastery, and accomplishment, effective tool use may coexist with diminished professional fulfillment. This distinction matters because the difficulty concerns the significance of the work being transformed, not necessarily practitioners' ability to perform it. Technical proficiency, accountable control, and professional fulfillment are related but distinct aspects of adaptation.

This explanatory detail provides a basis for diagnosis. Governance ambiguity directs attention to permissible use and organizational guidance; burdensome verification directs attention to assurance practices and the resources allocated to them; disrupted identity and meaning direct attention to work design and opportunities to exercise valued expertise. Distinguishing these pathways helps identify which conditions an intervention needs to address and why a response appropriate to one cost may leave another unresolved. Although the effectiveness of such interventions requires evaluation, the model provides an empirical basis for identifying where changes to AI adoption and engineering work may reduce practitioners' psychological costs.

The diagnostic framework in Table~\ref{tab:diagnostic-framework} is an
analytical translation of our explanatory model. We derive the diagnostic
questions from the conditions associated with the emergence or intensification
of each psychological cost and the responses through which practitioners
attempt to manage it. Organizational conditions and work arrangements that
can be changed become intervention targets, from which we infer potential
responses. This translation connects the explanation of how costs arise to
the identification of where organizational action may be directed. During
AI adoption strategy design, the framework can help practitioners and managers examine anticipated demands and plan appropriate support. Throughout change management, it can guide recurring assessment of emerging costs, practitioners'
adaptations, and whether organizational responses alleviate costs or introduce
additional demands. The framework and proposed responses are implications
of the analysis whose practical effectiveness requires evaluation.

\subsection{Theoretical Implications}

Theoretically, we interpret the findings through the lenses of Conservation of Resources (COR) theory~\cite{hobfoll1989conservation,hobfoll2011conservation} and Self-Determination Theory (SDT)~\cite{deci2012self}. Although neither COR nor SDT guided the coding or analysis, both theories offer explanatory power for interpreting the psychological costs that emerged inductively. Accordingly, we use them to position our findings within existing theoretical frameworks, examine their alignment with these theories, and identify how our findings extend current theoretical understanding.

\begin{landscape}
\begin{table}
\centering

\caption{Psychological costs of AI adoption in SE and correspondence with technostress and additional explanatory value from our study.}

\label{tab:contribution}
\begingroup
\footnotesize
\renewcommand{\arraystretch}{1.12}
\setlength{\tabcolsep}{5pt}
\begin{tabularx}{\linewidth}{@{}
    >{\raggedright\arraybackslash}p{0.15\linewidth}
    >{\raggedright\arraybackslash}p{0.20\linewidth}
    >{\raggedright\arraybackslash}X@{}}
\toprule
\textbf{Psychological cost} &
\textbf{Closest technostress correspondence} &
\textbf{Additional explanatory value for AI adoption in SE} \\
\midrule

\textbf{Uncertainty distress} &
Techno-uncertainty; partial overlap with techno-complexity and
techno-insecurity~\cite{tarafdar2007impact,ragu2008consequences}. &
Distress does not necessarily arise from insecurity or perceiving AI itself as
threatening. Our findings show that practitioners must establish workable practices and understand their changing professional role while technological capabilities evolve rapidly, actionable knowledge remains limited, and organizational expectations
press for further adoption. \textbf{\emph{The model explains how these conditions jointly make adaptation psychologically demanding.}} Beyond identifying
techno-uncertainty, \textbf{\emph{it locates the cost in the tension between expectations to progress and the conditions needed to establish stable practice.}} This makes organizational expectations and learning conditions relevant intervention
targets alongside technological change. \\

\addlinespace[0.65em]

\textbf{Accountability anxiety} &
Role ambiguity and technology-related strain; partial
correspondence~\cite{ayyagari2011technostress,jeyam2026still}. &
Anxiety can arise even when responsibility is unambiguous; practitioners know
they remain accountable but cannot confidently understand, validate, or justify
the outputs for which they must answer. \textbf{\emph{Our model identifies the gap between retained accountability and the perceived ability to discharge it, with
governance ambiguity and established quality obligations intensifying that
gap.}} Beyond identifying role ambiguity or technology-related strain, this
explanation shows why clarifying who is responsible may be insufficient. Practitioners also need conditions that enable them to exercise that
responsibility. \\

\addlinespace[0.65em]

\textbf{Cognitive load intensification} &
Techno-overload and techno-complexity, and associated
strain~\cite{tarafdar2007impact,ragu2008consequences}, partial correspondence. &
Cognitive demands arise partly from practitioners' efforts to manage
accountability anxiety. Verification requires them to reconstruct sufficient
understanding of delegated work to assess and endorse it. \textbf{\emph{Our model therefore explains how mechanisms used to verify AI-generated work create additional cognitive demands; practitioners must understand the generated solution, assess its correctness, and justify accepting it.}} This directs attention to the scope of delegated work and the time and support required for meaningful verification. \\

\addlinespace[0.65em]

\textbf{Craft identity disruption} &
Techno-insecurity; partial correspondence through threats to expertise and
professional standing~\cite{ragu2008consequences}. &
The cost concerns continuity of professional identity, extending beyond fears
of replacement or skill obsolescence. Practitioners may retain valuable
expertise while losing opportunities to enact it through activities that
previously defined their professional selves. \textbf{\emph{Our model connects AI's
substitution of these activities to the effort required to reconstruct
professional identity.}} Beyond identifying techno-insecurity, it distinguishes
possessing competence from being able to express that competence in
professionally meaningful ways. This may explain why training or employment
reassurance may leave the underlying identity disruption unresolved. \\

\addlinespace[0.65em]

\textbf{Meaning and satisfaction erosion} &
Reduced job satisfaction: an outcome rather than a technostress
creator~\cite{ragu2008consequences}. &
Reduced satisfaction can arise because AI assumes activities that previously
provided enjoyment, ownership, and accomplishment, even when practitioners
recognize its usefulness. \textbf{\emph{Our model connects changes in the composition of engineering work to changes in its intrinsic rewards. Beyond identifying dissatisfaction as an adverse outcome, it explains what has become less
rewarding and why.}} This distinguishes reducing strain from restoring
fulfillment; making AI-assisted work easier or less stressful does not
necessarily restore the meaning derived from the activities it replaces. \\

\bottomrule
\end{tabularx}

\par\smallskip
\parbox{\linewidth}{\footnotesize\emph{Note:} Correspondences identify related
stressors or outcomes rather than equivalent constructs. Proposed intervention
points require empirical evaluation; they are not prescribed recommendations but opportunity for reflection.}
\endgroup

\end{table}
\end{landscape}

\begin{landscape}
\begin{table}[]

\centering
\caption{A diagnostic framework for identifying intervention points during AI adoption in software engineering.}
\label{tab:diagnostic-framework}
\begingroup
\small
\renewcommand{\arraystretch}{1.12}
\setlength{\tabcolsep}{5pt}

\begin{tabularx}{\linewidth}{@{}
    >{\raggedright\arraybackslash}p{0.14\linewidth}
    >{\raggedright\arraybackslash}X
    >{\raggedright\arraybackslash}p{0.17\linewidth}
    >{\raggedright\arraybackslash}X@{}}
\toprule
\textbf{Psychological cost} &
\textbf{Diagnostic questions} &
\textbf{Intervention target} &
\textbf{Potential response to evaluate} \\

\midrule
\addlinespace[0.65em]

\textbf{Uncertainty distress} &
Can practitioners identify useful AI applications for their role and establish
workable learning priorities? Do adoption expectations account for
technological instability and uneven applicability? &
Learning conditions and adoption expectations &
Provide protected learning time and guidance relevant to particular roles.
Calibrate adoption expectations to applicable use cases, compliance
constraints, and delivery obligations. \\
\addlinespace[0.85em]

\textbf{Accountability anxiety} &
Are practitioners expected to approve outputs they cannot adequately
understand or validate? Are permissible use, assurance responsibilities, and
escalation routes sufficiently clear? &
Governance and assurance arrangements &
Clarify permissible use and assurance responsibilities. Establish review
criteria, access to appropriate expertise, and escalation routes for
unresolved quality or compliance concerns. \\
\addlinespace[0.85em]

\textbf{Cognitive load intensification} &
What cognitive effort is required to understand, verify, and correct generated
outputs? Is this effort recognized in workload and productivity expectations? &
Work allocation and scope of delegation &
Allocate capacity for verification. Bound delegated tasks and generated
changes so they remain reviewable. Assess the combined effort of generation,
evaluation, and correction. \\
\addlinespace[0.85em]

\textbf{Craft identity disruption} &
Which activities enable practitioners to express professional expertise and
ownership? What opportunities remain to exercise these capabilities as AI
assumes parts of the work? What psychological and professional support do practitioners need during adoption? What career advices are available to practitioners?&
Professional agency and role design &
Involve practitioners in redesigning their roles. Preserve meaningful
decision authority and opportunities to exercise, develop, and demonstrate
valued expertise. \\
\addlinespace[0.85em]

\textbf{Meaning and satisfaction erosion} &
Which activities provide interest, accomplishment, and enjoyment when AI replaces traditional work? How has AI changed access to these activities, and does the revised work provide comparable sources of fulfillment? &
Task composition and sources of fulfillment &
Review the balance of creative, routine, craftsmanship values, and evaluative work. Preserve opportunities for valued problem solving and assess fulfillment alongside
productivity. \\
\addlinespace[0.65em]

\bottomrule
\end{tabularx}

\par\smallskip
\parbox{\linewidth}{\small\emph{Note:} Apply the questions to practitioners'
specific tasks and working conditions. Costs and intervention targets may
overlap.}
\endgroup

\end{table}
\end{landscape}

While COR theory explains how individuals experience stress when valued psychological resources are threatened, depleted, or require continued investment to protect them~\cite{hobfoll1989conservation}, SDT explains how the satisfaction of basic psychological needs shapes motivation, well-being, and functioning of individuals~\cite{deci2012self}. Both theories provide complementary interpretations for understanding how organizational AI adoption both threatens practitioners' psychological resources and alters the motivational foundations of their work.

\begin{figure*}[th!]
    \centering
    \includegraphics[trim=2cm 4cm 0cm 1cm,clip,width=\textwidth]{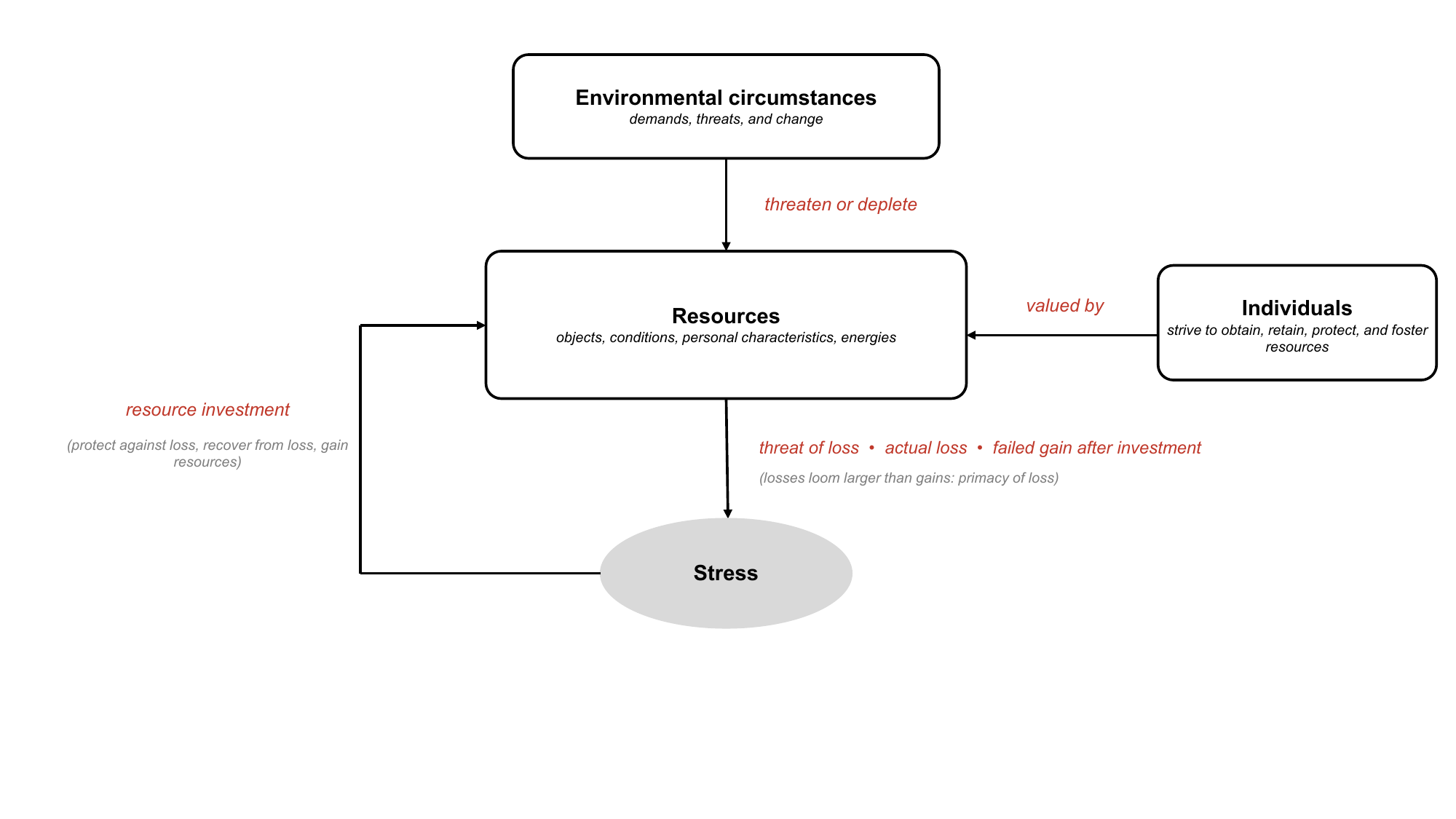}
    \caption{An aggregated view of Conservation of Resources theory~\cite{hobfoll1989conservation}.}
    \label{fig:COR_theory}
\end{figure*}

\paragraph*{\textbf{Conservation of Resources Theory}} COR's core concept is that ``stress'' occurs when ``people strive to retain, protect, and build resources and that what is threatening to them is the potential or actual loss of these valued resources''~\cite{hobfoll1989conservation}. Our definition of psychological costs is compatible with COR because the adverse cognitive, emotional, motivational, and social experiences identified in our findings (e.g., accountability anxiety) consistently emerged when practitioners perceived valued psychological resources (e.g., agency) to be threatened, depleted, or requiring sustained investment to protect. In this sense, COR does not redefine the psychological costs identified in our study; rather, it provides a theoretical explanation for why these experiences arise.

As depicted by Fig.~\ref{fig:COR_theory}, COR theory posits that individuals are fundamentally motivated to obtain, retain, protect, and foster the resources they value, encompassing objects, conditions, personal characteristics, and energies. Psychological stress occurs when environmental circumstances, such as demands, threats, or change--threaten or deplete these resources. Specifically, stress is triggered by the threat of loss, actual loss, or a failure to gain resources after investment, with resource losses being disproportionately more impactful than equivalent gains (the primacy of loss). In response to this stress, individuals must engage in resource investment to protect against further depletion, recover from existing losses, and ultimately build new resources~\cite{hobfoll1989conservation}.

Table~\ref{tab:cor-mapping} documents the mapping of COR theory constructs to our study's findings. Our conceptual model aligns closely with this foundational framework. Specifically, the dual forces of AI disruption and organizational adoption represent the environmental circumstances and demands that alter the practitioners' work environment. The five psychological costs we identified operationalize the stress that occurs when these environmental demands threaten or deplete practitioners' cognitive energies, professional conditions, and personal characteristics. In response, practitioners engage in resource investment through adaptive practices (managing, mitigating, and absorbing) aimed at protecting and recovering their most valued psychological resources: agency, competence, and professional identity.

This study makes four key theoretical contributions, to COR in the context of SE, by mapping the psychological costs of AI adoption to Conservation of Resources (COR) theory. First, we extend COR's concept of environmental stressors by modeling AI adoption as a dual-force disruption, where intrinsic technological uncertainties are actively amplified by extrinsic organizational pressures. Second, we specify the precise valued resources threatened in AI-mediated knowledge work, identifying the depletion of agency, competence, and professional identity as the core mechanisms of psychological cost. Third, we introduce agency displacement as a critical boundary condition, demonstrating that the severity of resource loss is contingent upon the extent to which AI substitutes a role's core generative activities. Finally, we expand COR's resource investment typology by identifying absorption as a distinct adaptive response, revealing that when resource loss (e.g., to professional identity) is perceived as irreversible, practitioners cope through resigned accommodation rather than active resource recovery.

\paragraph*{\textbf{Self-Determination Theory}} Self-Determination Theory (SDT) explains human motivation and well-being through the satisfaction of three basic psychological needs: autonomy, competence, and relatedness~\cite{deci2012self}. According to SDT, these needs are innate psychological nutriments that sustain psychological growth, integrity, and well-being, and their frustration is associated with diminished motivation and psychological functioning~\cite{deci2012self}.

Autonomy refers to experiencing one's actions as self-endorsed and congruent with one's values and sense of self~\cite{deci2012self}. SDT proposes that supporting autonomy fosters intrinsic motivation, enhances performance, and promotes psychological well-being~\cite{deci2012self}. Competence refers to the need to feel capable of effectively interacting with and influencing one's environment, and achieving valued outcomes through one's own abilities~\cite{deci2012self}. SDT argues that opportunities to develop and demonstrate competence are fundamental to intrinsic motivation and sustained engagement~\cite{deci2012self}. Relatedness refers to the need to feel connected to, valued by, and supported by others~\cite{deci2012self}. According to SDT, intrinsic motivation and psychological well-being are more likely to flourish in environments that foster secure interpersonal relationships and a sense of belonging~\cite{deci2012self}.

Table~\ref{tab:sdt-mapping} suggests that our findings align closely with the central propositions of SDT~\cite{deci2012self}. The psychological costs experienced during organizational AI adoption can be interpreted as reflecting the frustration of practitioners' basic psychological needs, mainly autonomy and competence. Autonomy was undermined when practitioners no longer experienced themselves as the originators of their work or were unable to confidently stand behind AI-assisted work and decisions. Competence was frustrated by the rapid evolution of AI, the continual obsolescence of emerging expertise, and the additional cognitive effort required to verify AI-generated outputs. By comparison, evidence relating to relatedness was less pronounced. While governance ambiguity and role-specific uncertainty sometimes weakened practitioners' sense of organizational support, the dominant experiences observed in this study concerned autonomy and competence rather than interpersonal connectedness.

Our findings demonstrate SDT's explanatory value in the context of organizational AI adoption in SE. We identified concrete software engineering practices through which autonomy and competence become frustrated during AI-assisted work, including diminished authorship, accountability without explainability, perishable expertise, and concerns over skill atrophy. Furthermore, the response patterns observed in our study, such as manual coding rituals, staged AI delegation, and identity reframing, can be interpreted as practitioners' attempts to re-establish autonomy and competence while adapting to AI-assisted software engineering. Our findings also broaden SDT application by showing how autonomy and competence are renegotiated during organizational AI adoption.

\begin{landscape}

\begin{table}

\caption{Mapping of Psychological Costs of AI Adoption to Conservation of Resources (COR) Theory.}

\label{tab:cor-mapping}
\renewcommand{\arraystretch}{1.8}
\footnotesize
\centering

\begin{tabular}{p{4.5cm}p{5cm}p{5.5cm}p{4cm}}
\toprule
\textbf{COR Theory Construct} & \textbf{Study Finding} & \textbf{Manifestation in AI Adoption} & \textbf{Resource Type Affected} \\
\midrule

\multirow{3}{=}{\textbf{Environmental Circumstances} \\ (demands, threats, change)} 
& AI Disruption 
& Fast technological pace, unknown risks, role-altering capability 
& \multirow{3}{=}{Conditions, Personal characteristics, Energies} \\
\cmidrule(lr){2-3}
& Organizational AI Adoption 
& Perceived adoption pressure, governance ambiguity, epistemic uncertainty, pre-existing standards 
& \\
\midrule

\multirow{5}{=}{\textbf{Stress} \\ (threat of loss, actual loss, failed gain)} 
& Uncertainty distress 
& Threat of skill obsolescence, inability to form stable expectations 
& \multirow{5}{=}{Personal characteristics, Energies, Conditions} \\
\cmidrule(lr){2-3}
& Accountability anxiety 
& Actual loss of control over work products, threat to professional judgment 
& \\
\cmidrule(lr){2-3}
& Cognitive load intensification 
& Depletion of mental resources required for verification 
& Energies \\
\cmidrule(lr){2-3}
& Craft identity disruption 
& Actual loss of authorship and generative work 
& Personal characteristics, Conditions \\
\cmidrule(lr){2-3}
& Meaning and satisfaction erosion 
& Failed gain after investment, loss of intrinsic rewards 
& Personal characteristics, Conditions \\
\midrule

\multirow{3}{=}{\textbf{Resource Investment} \\ (protect, recover, gain)} 
& Manage responses 
& Oversight practices, staged delegation, verification 
& \multirow{3}{=}{Energies, Personal characteristics} \\
\cmidrule(lr){2-3}
& Mitigate responses 
& Risk containment, skill preservation, identity reframing 
& \\
\cmidrule(lr){2-3}
& Absorb responses 
& Resigned adaptation, acceptance of losses 
& \\
\midrule

\textbf{Resource Conservation Goal} 
& Preservation of psychological resources 
& Agency (control), Competence (capability), Professional Identity 
& Objects, Conditions, Personal characteristics \\
\bottomrule

\end{tabular}
\end{table}

\end{landscape}

\begin{landscape}
\begin{table}

\caption{Mapping of Psychological Costs and Responses to Self-Determination Theory (SDT).}

\label{tab:sdt-mapping}
\centering
\renewcommand{\arraystretch}{1.8}
\footnotesize

\begin{tabular}{p{4cm}p{5.5cm}p{7.5cm}}
\toprule
\textbf{SDT Construct} & \textbf{Study Finding} & \textbf{Interpretation in the Lens of SDT} \\
\midrule

\multirow{3}{=}{\textbf{Autonomy} \\ \textit{(The need to feel volitional, in control, and the originator of one's actions)}} 
& Accountability anxiety 
& AI's opaque reasoning thwarts autonomy by stripping practitioners of the ability to fully understand, justify, or self-endorse the work they are held accountable for. \\
\cmidrule(lr){2-3}

& Craft identity disruption 
& The shift from ``author'' to ``evaluator'' undermines the sense of personal authorship and volition, making practitioners feel like ``prompt monkeys'' rather than originators of their work. \\
\cmidrule(lr){2-3}

& Meaning and satisfaction erosion 
& When generative work is delegated to AI, practitioners lose the intrinsic motivation derived from self-directed creation, leading to a sense of alienation from their own work. \\
\midrule

\multirow{3}{=}{\textbf{Competence} \\ \textit{(The need to feel effective, capable, and able to master one's environment)}} 
& Uncertainty distress 
& The perishable nature of AI knowledge and rapid technological evolution thwart competence by continuously invalidating emerging routines before mastery can be achieved. \\
\cmidrule(lr){2-3}

& Cognitive load intensification 
& The heavy cognitive burden of line-by-line verification depletes mental energy, reducing the practitioner's perceived efficacy and capacity to perform their role effectively. \\
\cmidrule(lr){2-3}

& Fears of skill atrophy 
& Prolonged reliance on AI threatens the practitioner's belief in their long-term capability, creating anxiety that their core technical mastery will degrade over time. \\
\midrule

\multirow{2}{=}{\textbf{Relatedness} \\ \textit{(The need to feel connected, supported, and valued within a social context)}} 
& Governance ambiguity \& Adoption pressure 
& The organization's ``trust-by-default'' culture paradoxically isolates practitioners, transferring compliance liability onto the individual without providing supportive, clear guardrails or shared best practices. \\
\cmidrule(lr){2-3}

& Epistemic uncertainty (e.g., Testers) 
& When organizational ``experiment-and-share'' strategies yield results irrelevant to specific roles, practitioners feel disconnected from the broader team's learning process and undervalued in their specific contributions. \\

\bottomrule

\end{tabular}
\end{table}

\end{landscape}

\subsection{Boundary Conditions of the Findings}

Different psychological costs may have different boundary conditions~\cite{yin2013validity}, reflecting the organizational setting, the stage of adoption, and the relationship between AI and engineering work. Our purpose is to qualify the analytical transferability of the proposed explanation~\cite{yin2013validity}. The conditions discussed below are informed by the case; their effects have not been independently established.

\textbf{The significance of activities assumed by AI.} Agency displacement, craft identity disruption, and meaning erosion may depend on how practitioners value the activities delegated to AI. These costs may be particularly salient when AI assumes activities through which practitioners express expertise, authorship, and accomplishment. These costs may be weaker when work carries little personal significance. The relevant boundary therefore concerns practitioners' relationships with particular activities, and not occupational roles or adoption levels alone. 

\textbf{Accountability relative to assurance capacity.} The relationship between delegated software production and accountability anxiety may depend on practitioners' capacity to discharge their retained responsibilities. At \textsc{SoftHouse}, established quality obligations and regulated domains made assurance particularly consequential. However, regulation is not necessarily a prerequisite; comparable tensions may arise wherever practitioners remain responsible for software they cannot adequately understand, validate, or justify. Where outputs remain reviewable and practitioners have sufficient expertise, time, and support, delegation may not produce the same anxiety or verification burden. This boundary concerns the relationship between responsibility and the conditions for exercising it.

\textbf{Adoption expectations relative to learning conditions.} The uncertainty pathway may depend partly on whether organizational expectations accommodate the difficulty of establishing workable practices. In our case, practitioners navigated rapid technological change, limited actionable knowledge, and uneven applicability while continuing to meet delivery obligations. These conditions may produce different experiences where adoption expectations reflect applicable use cases, compliance constraints, and available learning capacity. Technological uncertainty may persist in such settings, but the additional pressure to demonstrate progress need not take the same form. Accordingly, uncertainty distress should not be interpreted solely as insecurity or a perception of AI as threatening. Future developments in AI products may resolve some current limitations and make previously unsuitable use cases viable, reducing some sources of uncertainty while introducing new demands for learning and adaptation.

\textbf{Discretion over AI use.} Some responses identified in the model depend on practitioners retaining discretion over how work is performed. Bounded delegation, deliberate verification, and preserving opportunities for manual work require some influence over task allocation and working practices. AI adoption at \textsc{SoftHouse} was encouraged but not mandated, although organizational expectations still exerted pressure. Where AI use is tightly prescribed, these responses may be less available. The observed patterns of managing, mitigating, and absorbing costs should therefore not be assumed to transfer unchanged to those settings.

Where organizational prescriptions restrict practitioners' ability to limit delegation or retain manual work, they constrain practices through which costs were managed or mitigated in our case. For example, preserving competence through manual coding depends on having opportunities to continue that activity. Practitioners facing tighter restrictions may therefore require alternative responses. The categories of managing, mitigating, and absorbing costs may remain applicable, but the practices through which they are enacted and their relative prominence may differ. Our study does not establish how practitioners respond when these choices are constrained.

\textbf{The stage of adoption.} Our explanation reflects experiences approximately one year after the adoption's launch, while governance and working practices were still developing. As usable knowledge and assurance arrangements become more established, some sources of uncertainty may diminish. Meanwhile, further delegation may introduce different demands on verification, professional identity, and meaning. Our model therefore does not imply that costs inevitably decline with experience or increase with adoption. Our snapshot cannot establish which relationships persist, weaken, or change over time; these temporal boundaries require longitudinal investigation.

\subsection{Practical Implications \& Future Research}

While our findings identify a spectrum of psychological costs associated with organizational AI adoption, we focus our practical implications on two related implications: \emph{\textbf{agency displacement}}, and the \emph{\textbf{verification tax}}. We prioritize these themes for two reasons. First, they directly challenge the dominant industry narrative that frames generative AI predominantly as a productivity multiplier, revealing instead the hidden cognitive and professional costs of AI-supported SE work. Second, they explain the root cause of practitioner behaviours in AI-human collaboration other than ``resistance to change,'' but as a rational response to the psychological and professional challenges.

Our findings suggest that successful organizational AI adoption depends not only on deploying increasingly capable AI systems but also on understanding how AI redistributes human work and challenges identity in SE. While agency displacement draws attention to the extent to which AI assumes activities through which practitioners exercise professional judgment and professional identity, the verification tax highlights the additional cognitive and engineering effort required for practitioners to remain accountable for their work when AI-assisted.

\paragraph*{\textbf{Agency Displacement}}

We found that psychological costs are not uniform, but probably proportional to the degree of agency displacement, i.e., the extent to which AI substitutes the core generative activities through which practitioners exercise their expertise. The greater this displacement, the greater the psychological costs experienced by practitioners. Software engineers, whose professional identity is deeply tied to authoring code, experience the highest displacement. The psychological costs of AI adoption seem to be proportional to the degree to which AI replaces the core activities through which a practitioner exercises their professional agency. It shows that the psychological costs we observed maybe not resistance to change, but a rational and a human response to the displacement of the specific tasks that define a role's professional worth.

The Software Craftsmanship movement argues that software engineering is more than producing functional code; it is a professional craft emphasizing mastery, continuous learning, pride in workmanship, and responsibility for quality~\cite{craftsmanship2009}. Sundelin et al. demonstrates that professional identity in software engineering is deeply tied to specific generative activities, such as writing clean code, refactoring, test-focused development, and participatory architecture~\cite{sundelin2021towards}. Our study shows that when AI systems assume these core activities, they do not merely automate tasks; they can challenge professional identity, and established norms of craftsmanship. These findings help explain how AI adoption challenges professional identity and craftsmanship standards. \textbf{Implication:} Our findings suggest that successful organizational AI adoption should be evaluated not only by the extent to which AI augments software development, but also by the extent to which it preserves opportunities for practitioners to exercise meaningful professional agency and craftsmanship qualities.

Our findings also suggest that a single AI adoption strategy is unlikely to be equally effective across software engineering roles. Because the degree of agency displacement varies by role, organizations should tailor AI integration accordingly. \textbf{Implication:} AI adoption for software engineers, should give more attention to preserving opportunities to engage in creative and generative activities, whereas adoption strategies for architects or testers may focus more on augmenting decision making.

\textbf{Future Research:} Future work should develop agency displacement as an empirical construct by identifying measurable dimensions, validating instruments, and examining how different levels of displacement relate to psychological costs, job satisfaction, and long-term adoption outcomes across software engineering roles.

\paragraph*{\textbf{Verification Tax}}

For software engineers, AI adoption seems to shift accountability from production and quality to verification. While AI reduces the manual effort of writing code, it drastically increases the cognitive burden of verifying, justifying, and taking accountability for AI-generated outputs. AI introduces a significant ``verification tax'' by shifting software engineers' cognitive load from production to evaluation.

Conceptually, the verification tax does not represent a sixth psychological cost, but rather the direct synthesis of the accountability anxiety and cognitive load intensification identified in our findings. It captures the specific cognitive and professional toll exacted when practitioners are required to bridge the gap between their retained accountability for AI outputs and their diminished control over their generation. As they remain fully accountable for AI-generated outputs, they expend intense mental resources to verify, justify, and debug opaque code. If engineers cannot understand how the AI arrived at a solution, the cognitive load of verifying it becomes unsustainable.

Software engineers do not verify code because they finished writing it; they verify code while they are writing it. Software engineers are traditionally educated to solve problems through an integrated cognitive process in which understanding the problem, designing a solution, writing code, and verifying correctness evolve together. By externalizing code generation, AI compresses this process into predominantly post hoc verification, requiring practitioners to evaluate solutions without having participated fully in the reasoning that produced them. This raises important questions for AI adoption in software engineering, and education. \textbf{Implication:} In industry practice, the existence of a verification tax dictates that the productivity gains of AI cannot be measured solely by code generation speed. Organizations must explicitly account for the increased cognitive load of code verification in capacity planning and sprint estimates. 

\textbf{Future Research.} AI tool builders must prioritize verification and provenance-tracking features to reduce the mental burden of validating opaque outputs, ensuring that the engineer's shift toward evaluation remains a sustainable and safe engineering practice. 

Our findings also draw attention to software engineering education to preserve the fundamental competencies through which students develop the ability to reason about, construct, and verify software, as meaningful verification is only possible when engineers possess sufficient conceptual and technical understanding to critically evaluate AI-generated solutions. At the same time, AI-assisted development introduces a new class of verification skills, including evaluating AI reasoning, identifying subtle errors, assessing generated code against architectural intent, and deciding when AI outputs can or cannot be trusted, that deserve explicit attention in future curricula. \textbf{Implication:} Curricula needs to address ``AI-assisted verification,'' teaching them how to critically evaluate, debug, and trace the reasoning of machine-generated code. Furthermore, pedagogical approaches should seek to understand how manual coding rituals promote skills or how ``self-then-AI'' sequencing may assist students develop the foundational cognitive links between problem understanding and solution validation before relying on automation.

\textbf{Future Research.} Future educational research should therefore investigate how these emerging verification competencies can be developed without compromising the foundational skills upon which effective verification depends.

\section{Research Trustworthiness}
\label{sec:trust}

\paragraph{\textbf{Meaning Saturation}}

We established trustworthiness through meaning saturation~\cite{hennink2017code}. We monitored saturation throughout the iterative analysis by continuously comparing new interview excerpts and Second Cycle Pattern Codes meaning against the evolving codes in prior iterations. Data collection was considered sufficient when additional interviews no longer expanded the conceptual properties of the identified themes or their relationships within the emerging model~\cite{hennink2017code}. Saturation was assessed across software engineering roles rather than within individual participant groups to ensure that the conceptual model captured the breadth of practitioners' experiences across the organization. Aware of our design choice of treating software practitioners as a single cohort, we carried out a magnitude analysis (see Tbl.~\ref{tab:role-prevalence}) to understand the variation across roles.

Because saturation was assessed at the sample level, we do not claim meaning saturation within individual roles, particularly those represented by only three participants. Accordingly, the magnitude analysis is descriptive, and the observed role-based patterns are treated as exploratory rather than conclusive.

\paragraph{\textbf{Member Checking}}

Member checking was conducted to establish the reliability of our findings. We provided participants with our interpretations and sought their feedback~\cite{birt2016member,thomas2017feedback}, see Sect.~\ref{sec:member-checking}. 

\section{Limitations and Trade-offs}
\label{sec:limit}

We treated ``software practitioners'' as a single cohort when studying AI adoption. Although this design carried limitations, our contribution provides a nuanced, role-contingent results. This treatment also yielded findings, we would not have otherwise. For example, the definition of agency displacement might not have been possible if we had only focused on software engineers.

Our findings are necessarily situated in the organizational context of \textsc{SoftHouse}. Its regulated domains, strong quality and accountability culture, mature engineering practices, and experiment-and-share strategy may have shaped how psychological costs were experienced and which organizational conditions amplified them. We therefore do not claim that the prominence of individual costs or the observed role patterns will transfer unchanged to other organizations. However, context dependence is intrinsic to explanatory case research~\cite{yin2018case}. Our aim is analytical; by describing the case context and distinguishing AI-related sources from organizational amplifiers, we enable readers to assess the transferability of the proposed constructs and relationships to other settings.

Our sample is not balance on software roles; it is skewed towards software engineers. While this design choice reflects \textsc{SoftHouse} software practitioners population, it may have introduced an internal threat to validity. Some findings are role-specific, i.e., verification tax. We may have also missed the opportunity to reveal richer findings for the other roles, i.e., senior management, architects, and testers.

The study captures a cross-sectional ``snapshot'' (data collected at a single and specific point in time) of practitioners' experiences at a specific point in time (i.e., one year into the AI adoption). However, AI and its adoption are evolving; the technology and organizational policies are moving targets. Therefore, the psychological costs and the the practitioners' responses (Manage, Mitigate, Absorb) may likely evolve. A cross-sectional design cannot determine the long-term trajectory of these costs and responses. For example, it remains unknown whether practitioners who currently ``absorb'' the loss of generative joy will eventually experience long-term burnout or attrition, or whether the feared ``skill atrophy'' will actually materialize and impact code quality over a 3-to-5-year period. This limitation is an opportunity for future studies to track developers over a long period to observe how their psychological resources deplete or recover, and how their adaptive responses evolve as AI tools become more explainable and integrated.

Nineteen of the 21 participants had more than ten years of professional experience. This strengthened the study because experienced practitioners could compare AI-mediated work with well-established engineering practices, professional identities, and accountability norms. However, their over-representation may have foregrounded costs associated with disrupted expertise and craft identity. Less-experienced practitioners, whose skills and professional identities are still developing, may experience AI adoption differently.

We did not conduct a gender-based analysis to identify gender specific patterns in our data. This is due to low representation of women in our sample. Future research may consider a gender-base design to investigate AI adoption impact on software professionals.

\section{Conclusion}
\label{sec:conclusion}

Artificial intelligence is rapidly becoming an integral part of software engineering practice, yet organizational AI adoption has largely been viewed through the lenses of productivity, efficiency, and technical capability. This study shows that such a perspective imaybe not always complete. Through an in-depth case study of organizational AI adoption, we identified five psychological costs experienced by software practitioners: \emph{uncertainty distress}, \emph{accountability anxiety}, \emph{cognitive load intensification}, \emph{craft identity disruption}, and \emph{meaning and satisfaction erosion}. We further showed that these costs are shaped jointly by the disruptive characteristics of AI and the organizational conditions through which AI is introduced, and that practitioners respond by managing, mitigating, or absorbing these costs to preserve valued psychological resources.

Beyond identifying these psychological costs, our findings contribute two concepts that help explain the human experience of AI adoption in software engineering. \emph{Agency displacement} explains why psychological costs vary across software engineering roles, highlighting that the extent to which AI assumes the activities through which practitioners exercise their expertise influences how AI adoption is experienced. \emph{Verification tax} reveals that while AI reduces the effort required to generate software artifacts, it simultaneously increases the cognitive effort required to understand, justify, and remain accountable for them. Together, these concepts re-position AI for SE from what AI can generate to what software practitioners must continue to do to remain responsible engineers.

Our findings suggest that successful organizational AI adoption should not be evaluated by productivity gains or technical augmentation only. Rather, it should also consider how AI reshapes the human experience of software engineering, including practitioners' agency, competence, professional identity, and capacity to exercise responsible engineering judgment. By bringing psychological costs into the study of AI adoption, we hope this work contributes to a more human-centered understanding of how AI can augment software engineering without diminishing the role of the people who practice it.

\section*{Generative-AI use}

Paperpal\footnote{\href{https://edit.paperpal.com/}{https://edit.paperpal.com/}} was used throughout the writing process and solely for text improvements. The draft was produced independently by the first author. First, the ``Grammar'' feature of Paperpal was used to fix grammar issues and proofreading. Subsequently, the ``Rewrite'' feature with the ``Improve fluency'' option checked was used to enhance the final text. The re-generated text from the tool was adopted after the first author verified the edits and accepted or rejected the suggestions as deemed appropriate. Once the paper's text became available, the second and third authors reviewed the text without tooling assistance.

\begin{acks}

We are thankful to participating organization and to all interviewees who generously shared their time, experiences, and insights. This study would not have been possible without their openness, and willingness to reflect honestly on their AI adoption journey. This work is supported by the Innovation Fund Denmark for the project AI4SE1DK (4354-00006B).

\end{acks}

\bibliographystyle{ACM-Reference-Format}
\bibliography{references}

\end{document}